\documentclass[fleqn,usenatbib,useAMS]{mnras}

\usepackage{graphicx}	
\usepackage{amsmath}	
\usepackage{amssymb}	
\usepackage{multicol}        
\usepackage{bm}		
\usepackage{pdflscape}	
\usepackage{multirow}

\usepackage{graphicx}	
\usepackage{amsmath}	
\usepackage{amssymb}	

\newcommand{\FWBSB}{F-WBSB}
\newcommand{\WBS}{WBS}
\newcommand{\GBM}{GBM}
\newcommand{\Batse}{BATSE}
\newcommand{\BAT}{BAT}

\newcommand{\cps}{change-points}

\usepackage[T1]{fontenc}
\usepackage{ae,aecompl}

\usepackage{newtxtext,newtxmath}

\title[Detection of gamma-ray transients with binary segmentation]{Detection of gamma-ray transients with wild binary segmentation}

\author[S. Antier]{
S. Antier$^{1,2}$,
K. Barynova$^{2,3}$, 
P. Fryzlewicz$^{4}$,
C. Lachaud$^{1}$,
G. Marchal-Duval$^{2}$
\\
$^{1}$APC, Univ. Paris Diderot, CNRS/IN2P3, CEA/lrfu, Obs de Paris, Sorbonne Paris Cit\'e, France\\
$^{2}$LAL, Univ Paris-Sud, CNRS/IN2P3, Orsay, France\\
$^{3}$Department of Physics, Taras Shevchenko National University, 03022 Kiev, Ukraine\\
$^{4}$Department of Statistics, London School of Economics, Houghton Street, London WC2A 2AE, UK\\}
\date{Last updated 2019 July 17; in original form 2019 July 18}

\pubyear{2019}

\begin{document}
\label{firstpage}
\pagerange{\pageref{firstpage}--\pageref{lastpage}}
\maketitle

\begin{abstract}
In the context of time domain astronomy, we present an offline detection search of gamma-ray transients using a wild binary segmentation analysis called \FWBSB{} targeting both short and long gamma-ray bursts (GRBs) and covering the soft and hard gamma-ray bands. We use NASA Fermi/GBM archival data as a training and testing data set. This paper describes the analysis applied to the 12 NaI detectors of the Fermi/GBM instrument. This includes background removal, change-point detection that brackets the peaks of gamma-ray flares, the evaluation of significance for each individual GBM detector and the combination of the results among the detectors. We also explain the calibration of the $\sim$\,10 parameters present in the method using one week of archival data. Finally, we present our detection performance result for 60 days of a blind search analysis with \FWBSB{} by comparing to both the on-board and offline \GBM{} search as well as external events found by others surveys such as Swift-\BAT{}. We detect 42/44 on-board \GBM{} events but also other gamma-ray flares at a rate of 1 per hour in the 4-50 keV band. Our results show that \FWBSB{} is capable of recovering gamma-ray flares, including the detection of soft X-ray long transients. FWBSB offers an independent identification of GRBs in combination with methods for determining spectral and temporal properties of the transient as well as localization. This is particularly useful for increasing the GRB rate and that will help the joint detection with gravitational-wave events.
\end{abstract}




\section{Introduction}

The most recent LIGO/Virgo O2 observational campaign opened a new era of multi-messenger time-domain astronomy \citep{2018arXiv181112907T,2019arXiv190103310T} with the detection of a binary neutron star coalescence GW170817 on 2017 August 17 at 12:41:06 UT followed by a gamma-ray burst (GRB170817A) detected by the Fermi Gamma-ray Burst Monitor (GBM) \citep{2017ApJ...848L..13A}.
\GBM{} continuously observes the entire sky that is not occulted by the Earth in the band 4 keV - 40 MeV with a 2 microsecond timing resolution with its 12 NaI and 2 BGO detectors \citep{2009ApJ...702..791M}. On-board, \GBM{} continuous data searches have been developed to detect in real time GRBs at a rate of about 240 GRBs per year \citep{Bhat16}. These GRBs detections have contributed to different joint analyses, including with other gamma-ray surveys like Swift-BAT \citep{2018ApJ...862..152K} or Insight-HXMT \citep{2018RAA....18...57Z} but also to find gamma-ray counterpart of orphan afterglows detected by optical surveys \citep{2016MNRAS.455..712L} and for new optical facilities that will be operational in the next decade such as LSST and ZTF \citep{2018MNRAS.481.4355D}.
\smallbreak
In the past with \Batse{} \citep{Stern02} and more recently with \GBM{}, it has been demonstrated that the sample of gamma-ray bursts detected by the on-board triggering of the instrument is large, but it is still smaller than what it could be at its sensitivity (see \href{https://gcn.gsfc.nasa.gov/fermi_gbm_subthreshold.html}{GBM subthreshold analysis}). Identifying the weaker GRBs may give a substantial increase of the GRB statistics : it may extend the log N-log Peak Flux to low Peak-flux and thus allow for the estimate of the total rate of GRBs in the universe. More importantly nowadays, with the emergence of the multi-messenger astronomy with the first joint GRB-GW detection, new analyses from the GBM team have been developed to detect gamma-ray counterparts of transient events. As an example, there is the LB15 method to search the detectors' continuous data for short transient events in temporal coincidence with LIGO/Virgo compact binary coalescence triggers \citep{2015ApJS..217....8B,2019ApJ...871...90B}. Others  ground data analyses originate from Terrestrial Gamma-ray Flashes (TGFs) searches are looking for extra short GRBs in the GBM data \citep{2015EGUGA..17.9961B}. 
\smallbreak
In this paper, we propose a new method, \FWBSB{}, for an independent blind ground analysis of the 12 NaI GBM detectors data. \FWBSB{} \footnote{Fermi/GBM-like search with Wild Binary Segmentation for the detection of Bursts} is a search of weak and highly variable transient phenomena in gamma-rays as GRBs. We interpret the statistical problem of detecting GRBs against an otherwise constant background as the problem of change-point detection, in the sense that the start- and end-points of the GRB are estimated separately as change-points against the constant background.  This concept is not traditionally used by the standard methods of the GRB trigger literature \citep{2015ApJS..217....8B,2019ApJ...871...90B,2012A&A...541A.122S}. Non standard approachs (but not similar to our change-point method) have been investigated \citep{1998ApJ...504..405S,2013ApJ...764..167S}. They use bayesian blocks algorithms for  identifying structured variations similar to GRB profiles \citep{2005ApJ...627..324N,2006ApJ...643..266N,2011ApJ...740..104H,2014AAS...22335211H}. Different approach can target different populations of GRBs. Indeed, the category of transient sources detected by an instrument depends not only on its design properties (detector surface, energy band, field of view), but also on the transient algorithm architecture.
Besides, \FWBSB{} allows us to maintain a low computational complexity of the resulting method: change-point detection can typically be achieved in close to linear computational time in the number of data points, whereas a typical one-stage method for estimating both the start- and the end-point of a GRBs in a single pass through the data would need at least a
quadratic computational complexity. Our approach is therefore computationally fast. It also preserves a linear high temporal resolution (e.g 128 ms) for the length of the sequence defined by start-end stops change-points which is not the actual case with classical methods using a $\{0.128,0.256,0.512,1.024,.. s\}$ multi-scale resolution. Different methods of change-points have been proposed in the litterature (see as example \citealt{YAO1988181,killick}). Here we perform our change-point analysis using the Wild Binary Segmentation (WBS) method of \citet{Fryzlewicz14}, in light of its encouraging empirical performance.  In addition, we use median smoothing for the evaluation of the gamma-ra background allowing flexibility to analyse longer transients in the count rate compared to previous methods in the astrophysical field as well as exploring the softer X-ray band below 50 keV. Finally, we conduct our analysis with a set of about 9 parameters, orthogonal to hundreds of parameters used in existing methods.
In this paper, the \FWBSB{} procedure has been tested on 60 days of \GBM{} daily records of available detector and energy bands. Then, we have compared our gamma-ray candidates with existing catalogs such as \href{https://gcn.gsfc.nasa.gov/gcn/fermi_gbm_subthresh_archive.html}{onboard Fermi-GBM, subthresholds}, \href{https://swift.gsfc.nasa.gov/archive/grb_table/}{Swift-BAT}, \href{http://ibas.iasf-milano.inaf.it/}{INTEGRAL-IBAS}, \href{http://astrosat.iucaa.in/czti/?q=grb}{Astrosat-CZTI}, \href{https://gcn.gsfc.nasa.gov/konus_grbs.html#tc1}{Konus-Wind} and \href{https://gcn.gsfc.nasa.gov/maxi_grbs.html}{MAXI}. This work is a first step toward detecting candidates of gamma-ray bursts or other astrophysical sources from background. First, the multiple independent detection among the 12 \GBM{} detectors enables a low rate of cosmics. Secondly, the use of several energy bands gives a first idea of the nature of the transients (most of the X-ray galactic flares are detected below 50 keV). Finally the comparison with other catalog surveys demonstrates our capacity to find real events not detected by the Fermi GBM standard method. Our method aims to be integrated in systematic offline analyses for current missions such as GBM, SPI-ACS or HXMT but also for future mission such as SVOM \citep{2016arXiv161006892W}. Jointly with a localization approach as shown in \citet{2019ApJ...873...60B} or similar \citep{2015EGUGA..17.9961B}, the method can be used for detecting new GRBs especially multi-peak and softer ones.
\smallbreak
The paper is organized as follows. In Section 2, we describe the \GBM{} data used to calibrate and test the method. In Section 3, we describe the \FWBSB{} general methodology and outline the \WBS{} technique in more detail. In Section 4, we show the results of the calibration and parameters used for \FWBSB{}. In Section 5, we give our results concerning the detection of gamma-ray transients over 60 days of \GBM{} data and compare with others surveys. 

\section{Input Data}

We evaluate \FWBSB{} detection performance using daily records from the twelve semi-directional sodium iodide (NaI) detectors of the GBM{} space instrument, covering an energy range of 4 - 5000 keV. We use the continuous Time-Tagged Event data (TTE), which gives a list of photons in each detector with time and energy channel information (128 energy resolution). We then build time series with a temporal resolution of 128 ms and 4 energy channels [4-50], [50-100], [50-900] and [4-900] keV. 
\smallbreak
\GBM{} time series can be described as $\{ X_t\}_{t \, \in \left [1,T \right ]}$, with recorded events in one NaI detector, with $N$ \cps{} where locations $\eta_1,...,\eta_N$ correspond to high variations of the signal due to presence of variable or transient astrophysical sources.
The canonical model of the signal is of the form :
\begin{equation}
 \forall t \, \in \left [1,T \right ], \, \, X_t = f_t + b_t
\end{equation}
where $f_t$ is a deterministic, one-dimensional, piecewise-constant signal with \cps{}, and the sequence $b_t$ is the background noise with smoothly-variation expectation (not related to any variable or transient astrophysical event, see Section~\ref{bkgtreatment} for details).

\section{\FWBSB{} Pipeline}
\begin{figure*}
\begin{center}
\includegraphics[scale=0.5]{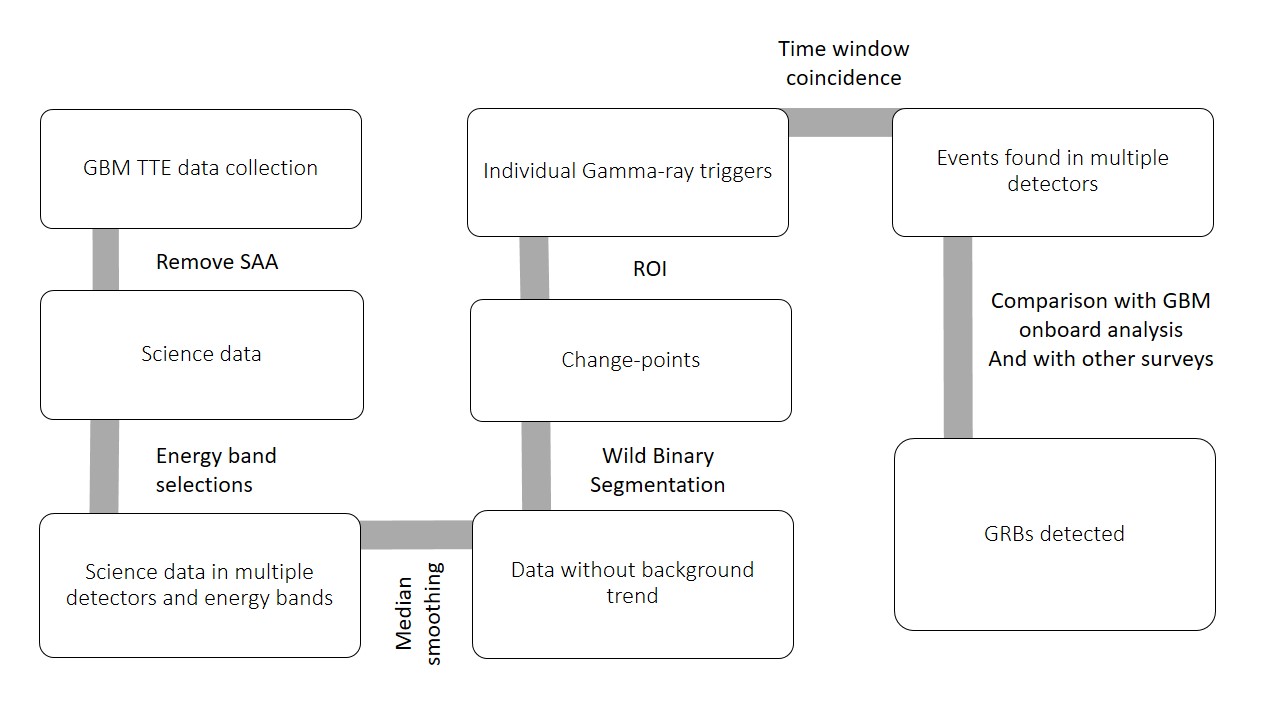}
\caption{Flow chart of the \FWBSB{} procedure. First, we collect the raw data, and we remove portions of the daily sequence corresponding to the South Atlantic Anomaly. Then we compute various light curves in different energy bands from the raw data for the different NaI detectors. We estimate the background with median smoothing to have a flat variation of the time series. We investigate the change-point detection with the \WBS{} approach \citep{Fryzlewicz14}. We construct different individual regions of interest (ROIs) and cross-match them in time to have obtain multiple-detector events. \label{fig:WBSflowchart}}
\end{center}
\end{figure*} 

\FWBSB{} is a multi-resolution timing approach for searching for transient variation lasting from 128\,ms to 50\,s in the four energy bands mentioned above to target both short and long GRBs. We now list the stages of the method as shown in in Figure~\ref{fig:WBSflowchart}:
\begin{itemize}
 \item \textbf{step 1 - science data}. We select portions of the \GBM{} time series that are suitable to be analyzed: e.g outside the South Atlantic Anomaly for which detectors are turned off to preserve their lifetime (see Section~\ref{science data}).

 \item \textbf{step 2 - background estimation}. \FWBSB{} works on a flat time series trend.
For this reason, we have to estimate $E(b_t)$ and subtract the background trend out before looking for any prompt excess as shown in Section~\ref{bkgtreatment}. The background trend arises because of contaminants such as bright high-energy sources that come in and out of the wide field of view, in addition to location-dependent particle and Earth atmosphere effects.
 \item \textbf{step 3 - change-points}. We compute the number and locations of multiple \cps{} in the cleaned data (after background removal) as in Section~\ref{changepointdetec}. To this end, we use the \WBS{} method of \citet{Fryzlewicz14}.
 \WBS{} does not require the choice of a window or span parameter and does not lead to a significant increase in computational complexity. The stopping criterion of \WBS{} uses the \textit{strengthened Schwarz Information Criterion}, which offers very good practical performance for rare phenomena. 
  \item \textbf{step 4 - region of interest}. The list of \cps{} is processed into a list of region of timing interval interests (ROIs), whose start and stop are the pair of \cps{} that maximize the signal to noise ratio of the ROI.
 \item \textbf{step 5 - multiple detections}. Individual-detector ROIs are cross-matched in time. A multiple-detector trigger e.g gamma-ray trigger is created with ends as the intersection of individual detector triggers. An evaluation of its significance measuring the excess over the background is performed in the contributed detector time series. 
 \item \textbf{step 6 - validation}. We proceed to an evaluation of the multiple-detector trigger reliability by comparing the different energy bands, the shape of the light curves, and a cross-match with onboard \GBM{} and untargeted search results \citep{2016arXiv161202395G} as well as other survey transients such as \BAT{} \citep{2016ApJ...829....7L}. 

\end{itemize}

 After this overview of the pipeline, we describe the details of each step.

\subsection{Science data}
\label{science data}
\begin{figure}
\begin{center}
\includegraphics[scale=0.30]{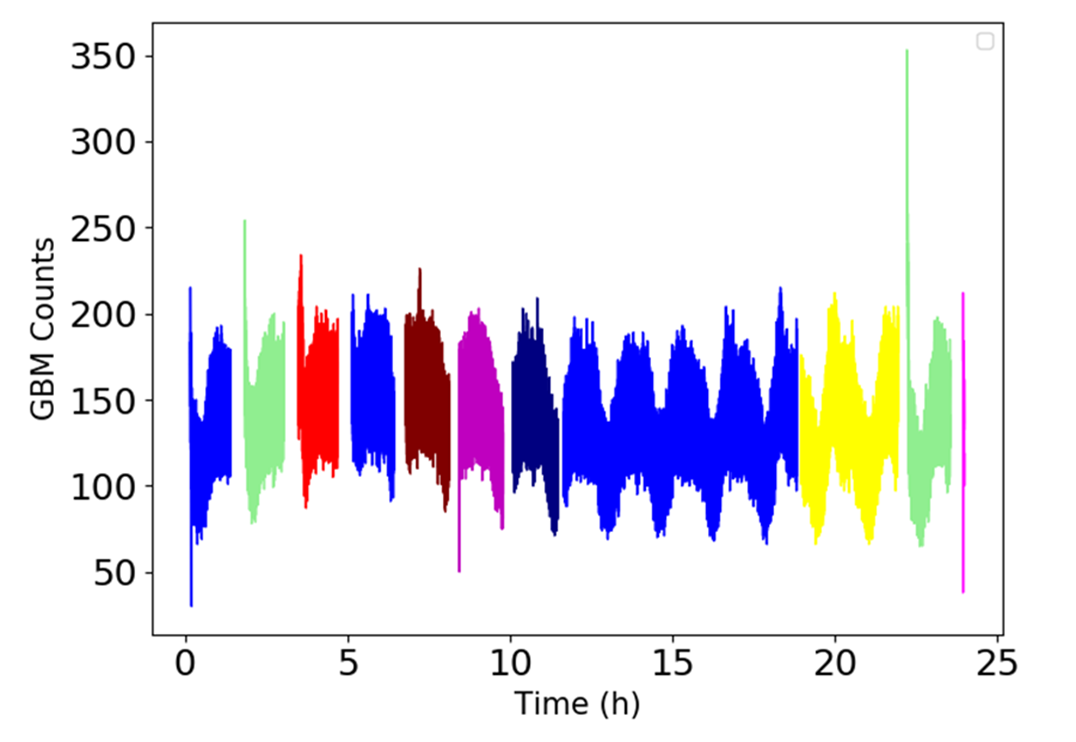}
\caption{\GBM{} counts detected (\GBM{} NaI $n_a$ detector) as a function of time (in hours) on September 4, 2018. Time bins are 0.128\,s wide. The counts of the energy bands (from 4 keV to 5 MeV) are summed. The different colors represent the portions of the light curve that will be analyzed (outside the South Atlantic Anomaly).}
\label{fig:SAA}
\end{center}
\end{figure}

We select portions of the \GBM{} time series outside the South Atlantic Anomaly (SAA) for which detectors are turned off and count rate is null (see Figure~\ref{fig:SAA}). We took a margin of 2.56\,s at each end. 

\subsection{Background estimation}
\label{bkgtreatment}
The \GBM{} detectors are subject to a substantial time-varying background, due to:
\begin{itemize}
 \item cosmic X-ray background as the highest contribution \citep{Moretti09},
 \item bright high-energy sources that come in and out of the wide field of view,
 \item South Atlantic Anomaly trapped particles,
 \item other particle flux,
 \item Earth transit in the field of view and Earth atmospheric effects.
\end{itemize}
We note that the different contributions to the background depend on the spectral window; moreover, the background does not have a specific profile timing variation.
In this analysis, the background is estimated for any time-interval (from milliseconds to dozen of minutes).
We use a median smoothing filter that offers the possibility to follow the trend of \GBM{} records. It is adaptable to any unexpected but relatively smooth variation of the background and differs from polynomial fitting approach in computing aspects and flexibility, and is not contaminated by high spikes of a GRB.

\smallbreak
\noindent The estimator $\tilde{E}(b_t)$ of $E(b_t)$ is of the form :
\begin{equation}
 \tilde{E}(b_t) =  {\widehat{\{ X_i \}}_{i  \in  [t  -  \ \frac{K-1}{2}, t\,  + \frac{K-1}{2} ]}} , \small \forall t  \in  [\frac{K-1}{2}+1,T-\frac{K-1}{2}  ] 
\label{Bandequ}
\end{equation}
with mirror-image padding at the left and right edge, where ${\widehat{\{ X_t\}}}$ is the median of $\{ X_t\}$ delimited by $K$, $K$ is a parameter to be defined, depending on the timing resolution of $\{ X_t\}$. For this analysis, $K$ value is shown in Table~\ref{tab:parameters}.

\begin{figure}
\begin{center}
\includegraphics[scale=0.36]{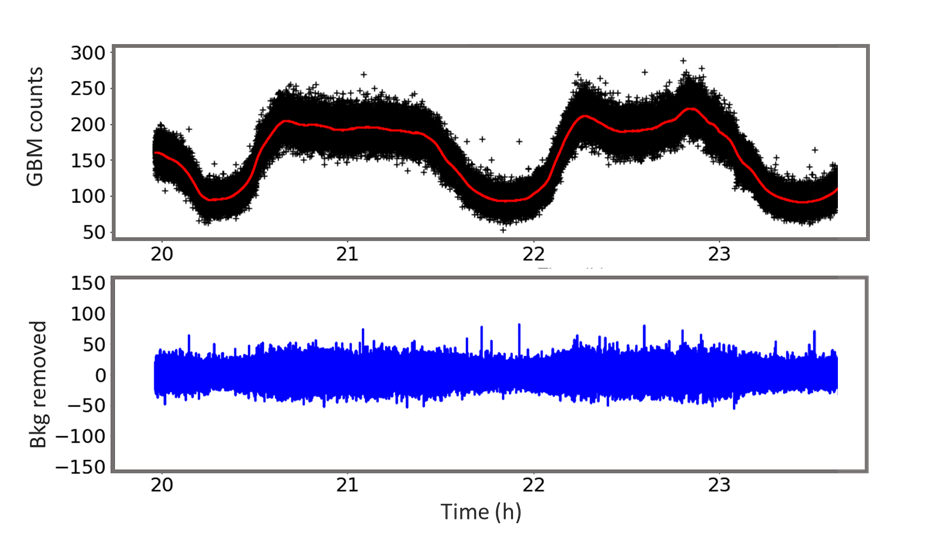}
\caption{a) \GBM{} counts detected (\GBM{} NaI $n_a$ detector) as a function of time (in hours) on September 1, 2018. Time bins are 0.128\,s wide. The counts of the energy bands (from 4 keV to 5 MeV) are summed. The red line represents the median smoothing estimate ($K=1025$, which corresponds to a median smoothing width of 130 s). b) Residual after background removal.}
\label{fig:mediansmoothing}
\end{center}
\end{figure}

\smallbreak
Figure~\ref{fig:mediansmoothing} represents the underlying background continuum of a \GBM{} detector, with 128\,ms principally dominated by the cosmic X-ray background (in black), with an estimate of the background trend using median smoothing (in red), and corresponding residual data (in blue). Note that the value of the span $K$ has been evaluated during the 7-day \GBM{} training session (see Table~\ref{tab:parameters} of Appendix \ref{Appendixextable}). 

\subsection{Change-point detection}
\label{changepointdetec}

\begin{figure}
\includegraphics[scale=0.53]{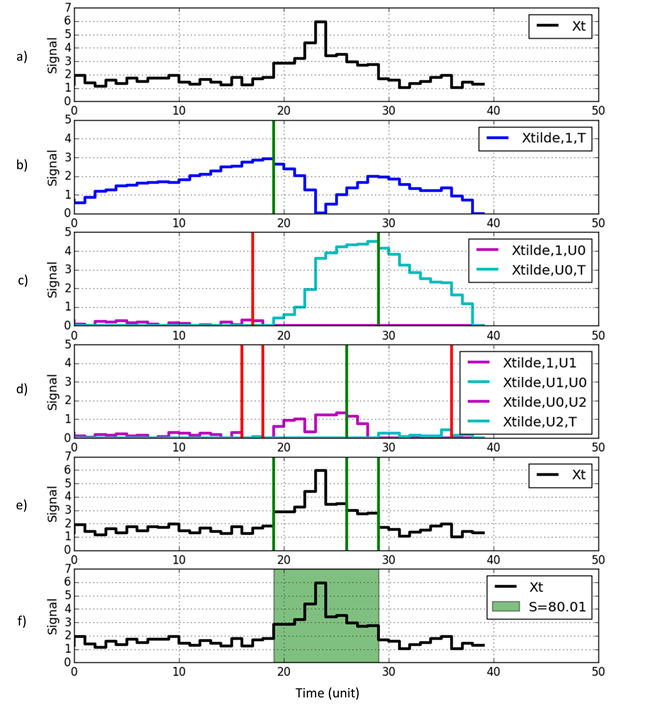}
\caption{An example of the \WBS{} iteration procedure. In steps a) to d), the change-point candidates
are iteratively determined by computing $|\tilde{X}_{s_m,e_m}^u|_{u \, \in [s_m,e_m]}$, where ${m \, \, \in \, \, \left [1, \,M \right ]}$. In this illustration, $M = 1$. They are then sorted from the most to the least important, and the N+1 most important ones, denoted by $\{ \tilde{\tau_i} \}_{i \, \in \left [0,N \right ]}$] and shown in green, are selected via the strengthened Schwarz Information Criterion. The rest, shown in red, are eliminated. Steps a) to d) use the \WBS{} algorithm. Step d) shows the set of detected change-points. In step e), the most significant regions of interest (ROI) are represented, delimited by two change-points. The selection of
contributed detectors will be done comparing the individual trigger times. Step f) shows the final ROI that is considered as a gamma-ray burst candidate.}
\label{fig:WBSmethod}
\end{figure}

\FWBSB{} searches for significant \cps{} in the count rate ($X_t$) recorded in \GBM{} detectors for different energy bands.
A pre-processing step has already removed the background trend (see Section \ref{bkgtreatment}).
The remaining and unexpected variations of the cleaned signal could be caused by:
\begin{itemize}
 \item a new transient source appearing in the field of view (the case we are interested in),
 \item high variability of X-ray sources present in the field of view,
 \item transit of X-ray sources on the edges of the field of view.
\end{itemize}

Change-points are determined by the \WBS{} procedure as described in \citet{Fryzlewicz14}.
Two \cps{} could reveal the start or the stop of a new transient source, constituting an interesting case for the purpose of this study. Below we list the steps of the method as illustrated in Figure~\ref{fig:WBSmethod}.


First, we define a set of $M$ random time intervals whose start- and end- points have been drawn (independently with replacement) uniformly from the set $\{1,T\}$.
\begin{eqnarray}
 \forall m \, \, \in \, \, \left [1,M \right ], & \, \, \, \, [s_m, e_m] \, \, \subseteq \, \, [1,T]  \nonumber \\
 \mathrm{If} \, \, M=1, & \,  \, \,\, \,\, \, \, \,\, \,\, \,\,  [s,e] \, \, = \, \, [1,T] 
\end{eqnarray}
\indent For each time interval $[s_m,e_m]_{m \, \, \in \, \, \left [1, \,M \right ]}$, we compute the \textit{contrast weights} vector. The inner product between this contrast vector and the vector ($X_{s_m},...,X_{e_m}$) is one of the basic ingredients of the wild binary segmentation algorithm. The \textit{contrast} vector $\tilde{X}_{s_m,\, e_m}^u$ given at the time $u$ is:

\begin{eqnarray}
\label{eq:ip}
\small \forall u \, \in \left [s_m,e_m \right ], \, \, n   =  e_m-s_m+1 \nonumber \,, & \\
\\ \nonumber
\tilde{X}_{s_m,\, e_m}^u  =  \sqrt{\frac{e_m-u}{n \,(u-s_m+1)}}\sum_{t=s_m}^u X_t - \sqrt{\frac{u-s_m+1}{n \, (e_m-u)}}\sum_{t=u+1}^{e_m} X_t \, .
\end{eqnarray}

We compute $(m_0, u_0) = \arg\max_{m, u \in \{s_m, \ldots, e_m-1\}} |\tilde{X}^u_{s_m,e_m}|$. We add $u_0$ to the set of change-point candidates. The time domain [1,T] is then split into two sub-intervals to the left and to the right of $u_0$ (see step b in Figure~\ref{fig:WBSmethod}). The recursion
continues in this way (see step c and d in Figure~\ref{fig:WBSmethod}) until all intervals $[s_m, e_m]$ have been examined (for the case $M>1$) or until there are no more intervals to consider (for the case $M=1$). We then order the change-points candidates from the most to the least important, according to the decreasing magnitude of $|\tilde{X}_{s_{m_0}, e_{m_0}}^{u_0}|$.
We determine which of the most important change-points enter the set of change-points $\{\tilde{\tau}_i\}_{i=0}^N$ (see step e in Figure~\ref{fig:WBSmethod}) through the so-called strengthened Schwarz Information Criterion, which is practically equivalent to the classical Schwarz (Bayesian) Information Criterion. If fewer than two change-points survive this selection, there are no burst candidates to speak of, and we stop. Otherwise, we next trigger the procedure
described in Section~\ref{Creation ROI}. Throughout the paper, we use $M = 5000$ shown in Table~\ref{tab:parameters}.
 
\begin{figure}
\begin{center}
\includegraphics[scale=0.32]{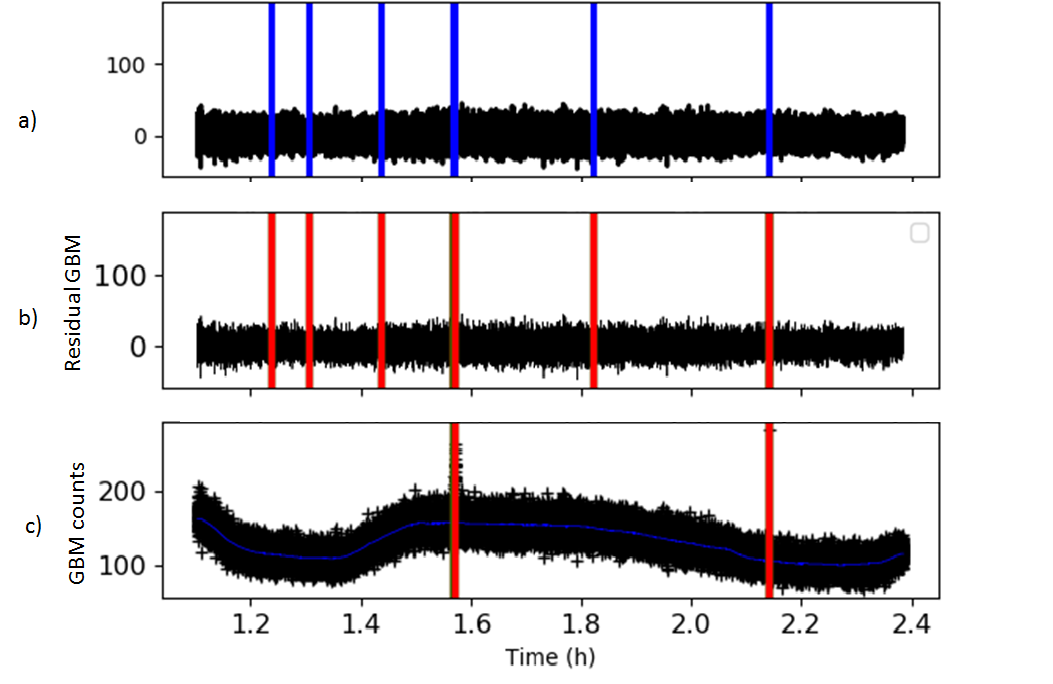}
\caption{Residual (case a and b after background removal) \GBM{} NaI $n_2$ detector counts as a function of time (in hours) on August, 08, 2017. Time bins are 0.128\,s wide. The counts of the energy bands (from 4 keV to 900 keV) are summed. a) blue vertical lines represent the change-points with \WBS{} b) green/red vertical lines represent list of ROI with start-and
end-points  (6 here) c) selection of the individual ROIs that pass the threshold $\mathrm{SNR}_\mathrm{max}$. Finally 2 ROIs selected.}
\label{fig:ROIchange}
\end{center}
\end{figure}

\subsection{Creation of the region of timing interest and evaluation of its significance}
\label{Creation ROI}

We evaluate the significance of a region of timing interest (ROI) with the signal to noise ratio (SNR) delimited by two pairs of change-points (not necessarily adjacent) with:
\begin{equation}
\mathrm{for} \, \mathrm{i,j} \, \in \left [0,N \right ], \, \, \mathrm{SNR_{\tilde{\tau_i},\tilde{\tau_j}}}=\frac{\sum_{t=\tilde{\tau_i}}^{t=\tilde{\tau_j}} \, X_t \, - \, \tilde{E}(b_t)}{\sqrt{\sum_{t=\tilde{\tau_i}}^{t=\tilde{\tau_j}}\tilde{E}(b_t)}}   
\end{equation} 
where $\left \{ \tilde{\tau_i} \right \}_{i \, \in \left [0,N \right ]} $ are change-point list found in previous step, $X_t$ GBM single detector time series, $\tilde{E}(b_t)$ background trend estimation.\\

If $[\tilde{\tau_i},\tilde{\tau_j}]_{i,j \, \in \left [0,N \right ]} \, \, < \, \, \Delta_{\tilde{\tau_i},\tilde{\tau_j}}$, and $\mathrm{SNR}_{\tilde{\tau_i},\tilde{\tau_j}} > \mathrm{SNR}_\mathrm{max}$, then the ROI is selected as significant for the next step. For this analysis, $\mathrm{SNR}_\mathrm{max}$ and $\Delta_{\tilde{\tau_i},\tilde{\tau_j}}$ values are shown in Table~\ref{tab:parameters}. Figure~\ref{fig:ROIchange} represent the residual (cases a and b) and (step c) \GBM{} daily data scan received in detectors ($n_2$) on September 8, 2008. The list of \cps{} detected are represented in case a) and the list of pairs of change points are listed in b). Finally, ROIs that passed the threshold $\mathrm{SNR}_{\mathrm{max}}$ are shown in c). Triggers around 1.56\,hrs correspond to GRB$170808065$ as listed in Table~\ref{tab:calGBMWBS} of Appendix \ref{Appendixextable}.
\smallbreak
We flag as ``1'' timing portions for which a single ROI has been found in a certain energy band using the process above. At the end of the process, if the value is above 2, it means it is flagged as a multiple-detector event. We re-calculate the signal to noise ratio in the joint timing window for the contributed detectors. We then compare the multiple-detector event to different GRB databases such as \GBM{} and \BAT{}.

\section{Calibration of \FWBSB{}}
\label{calibrationFwbsb}

We run the \FWBSB{} method over 7 \GBM{} daily data (from August 1 to August 7 2018 inclusive). Each data sequence is simultaneously analyzed for the NaI 12 count rates in the four energy bands mentioned.  We calibrate \FWBSB{} with the help of the 11 gamma-ray events found by GBM onboard. Our results are shown in section in Table~\ref{tab:calGBMWBS} of Appendix \ref{Appendixextable}. 
At the end of the calibration, all the parameters needed for this analysis are held constant (see Table~\ref{tab:parameters}). They are used for the 60-day \FWBSB{} performance described below.
\smallbreak
Results of the blind search show that \FWBSB{} detect 11/11 GRBs from the \href{https://gcn.gsfc.nasa.gov/fermi_grbs.html}{onboard GBM analysis} including long GRBs such as GRB180801492 and short GRBs such as GRB bn180801492. Note that \FWBSB{} found GRB180803590 in a single detection. We perform simulations for evaluating the false positive of the statistical background fluctuations. To do this, we create mock data using five full days (180808, 180809, 180907, 180810 and 180907). We compute a poissonian law distribution whose parameter was the median of any data science sequence in a given energy band. We run our \FWBSB{} analysis on the simulated data. We do not detect any multiple detection (two triggers found in the same time window of 60 s) for the different energy band. Our single detection rate (a detection in one detector) is at maximum 6 per day considering one energy band, leading a false positive rate of background fluctuations due to two detectors in coincidence at a level of 9 per year (with 60~s trigger time maximum separation.

\begin{table*}
\caption{Parameters from the \FWBSB{} method set-up after 7-days \GBM{} data investigation \label{tab:parameters}}
\begin{tabular}{cccc}
Name & Unit & Value & Description \\
\hline
Energy bands & keV & 4-50, 50-100, 50-900, 4-900 & \FWBSB{} searches are run independently.\\
Minimal resolution & ms & 128 & Minimal resolution for \FWBSB{} search \\
Kernel Median ($K$) & s & 130 & Kernel used for Background fitting for 128 ms resolution \\
 $M$ & & 5000 & random time intervals of \WBS{} \\
 $N$  & /hours & 3 & Number of maximum change-points accepted  \\
 $\Delta_{\tilde{\tau_i},\tilde{\tau_j}}$ & s & 50 & Maximum duration of \FWBSB{} duration \\
 $\mathrm{SNR}_\mathrm{max}$  & - & 3.0 & SNR threshold of a single detector trigger \\
\end{tabular}
\end{table*}

\section{Blind search of 60 \GBM{} daily records}

We run \FWBSB{} over 60 \GBM{} days with held constant intrinsic parameters as discussed in Section~\ref{calibrationFwbsb}: August 2017 (except 2017/08/20 and 2017/08/26, data not available), from 2018/08/08 to 2018/09/08. As shown in Table~\ref{tab:WBSstatistics}, \FWBSB{} detects more than 1000 events seen by at least two NaI detectors in 4-50 keV and 4-900 keV bands, and about 60 events in the 50-100 keV and 50-900 keV bands over this period. In parallel, we calculate the duty cycle of \GBM{} (e.g period when scientific analysis is done with exclusion of the South Atlantic Anomaly) as 46.5 days. Trigger rates lead to about 30 events per day for the 4-50 keV band, 20 events per day for the 4-900 keV band, 1.2 event per day in the 50-100 keV and 50-900 keV bands. All the results are available at \href{https://github.com/santier14/FWBSB/}{https://github.com/santier14/FWBSB/}. Events have different astrophysical origins, such as galactic flares, cosmic rays (mainly in the 4-50 keV band), variable stars or gamma-ray bursts. 

\begin{table}
\caption{\FWBSB{} results for the 60 days blind search analysis for August 2017 (except 2017/08/20 and 2017/08/26), from 2018/08/08 to 2018/09/08. The results are expressed in rates for all events found and for the joint \FWBSB{} and onboard GBM searches.  \label{tab:WBSstatistics}}
\begin{tabular}{ccccc}
Energy bands & \multicolumn{2}{c}{All triggers} & \multicolumn{2}{c}{GBM-WBS triggers} \\
& number & rate & number & rate \\
\hline
4-50 keV & 1508 & 32.40 /d & 32 & 0.67 /d \\
4-900 keV & 1049 & 22.54 /d & 39 & 0.84 /d \\
50-100 keV & 58 & 1.25 /d & 28 & 0.60 /d \\
50-900 keV & 57 & 1.22 /d & 35 & 0.75 /d \\
\end{tabular}
\end{table}

\subsection{Comparison with \GBM{} on-board analysis}

\FWBSB{} detects 42 gamma-ray bursts out of 44 GRBs (3/4 of short GRBs) found by the \GBM{} on-board analysis (see Table~\ref{tabresGBMWBS2} of Appendix~\ref{Appendixextable}). Most of the GRBs are seen in more than one energy band, but they represent 2/3 of the full \FWBSB{} events detected above 50 keV. Note that six GRBs are only recovered by a single detector event in the \FWBSB{} analysis. \href{https://heasarc.gsfc.nasa.gov/FTP/fermi/data/gbm/bursts/2017/bn170816258/quicklook/glg_lc_all_bn170816258.gif
}{GRB170816258} and \href{https://heasarc.gsfc.nasa.gov/FTP/fermi/data/gbm/bursts/2017/bn170817529/quicklook/glg_lc_all_bn170817529.gif
}{GRB170817529} are not detected by the blind \FWBSB{} search because no change-points were found at the GRB location. Nevertheless, we investigate locally without standard parameters,  10 minutes of data before the GRB170817529 trigger time until the next South Atlantic Anomaly passage happening a few minutes after GRB170817 reveals a multiple detection on n2 and n5 detectors with 1.152\,s and 0.640\,s durations respectively with a SNR of 6.1 in the 4 - 900 keV band.


\smallbreak

The number of detectors involved for the on-board \GBM{} analysis can vary in the \FWBSB{} analysis due to variations in the background fit during online and offline searches. As an example, \href{https://heasarc.gsfc.nasa.gov/FTP/fermi/data/gbm/bursts/2017/bn170830069/quicklook/glg_lc_all_bn170830069.gif}{GRB170830069} is triggered by $n_7$ and $n_b$ detectors with onboard GBM analysis whereas the GRB is clearly seen in $n_6$, $n_9$ and $n_a$ as WBS reported in Table~\ref{tabresGBMWBS2} of Appendix~\ref{Appendixextable}. 
\smallbreak

The \FWBSB{} and \GBM{} onboard analysis trigger times vary from less than 2\,s for more than 50\% of the 34 GRBs found in the 50-900 keV band. There is no clear preferential delay between the softer 4-50 keV band and the harder 50-900 keV band for the 32 GRBs found by \FWBSB{}. For example,  \href{https://heasarc.gsfc.nasa.gov/FTP/fermi/data/gbm/bursts/2017/bn170804911/quicklook/glg_lc_all_bn170804911.gif}{GRB170804911} is found 12\,s earlier by \FWBSB{} in the 4-50 keV band than \GBM{} and \FWBSB{} hard trigger analysis (above 50 keV). In contrast, \href{https://heasarc.gsfc.nasa.gov/FTP/fermi/data/gbm/bursts/2018/bn180906597/quicklook/glg_lc_all_bn180906597.gif}{GRB180906597} is found 2 s delay in the 4-50 keV compared to \FWBSB{} 50-900 keV search. 

Moreover, we also detect multi-episodes of 8 out of 9 long GRBs ($\mathrm{T90} \, >$ 100 s) which helps to recover its duration by identification. Indeed, \FWBSB{} configuration finds event/peak duration up to 50 s as mentioned in Table~\ref{tab:parameters}. Note \href{https://heasarc.gsfc.nasa.gov/FTP/fermi/data/gbm/bursts/2018/bn180826055/quicklook/glg_lc_all_bn180826055.gif}{GRB180826055} is detected by \FWBSB{} on detector $n_9$ at 01:17:22 e.g 123 seconds before the \GBM{} trigger, with a detection of a precursor gamma-ray emission. We compare the duration of the \FWBSB{} event duration in the 4-900 and 50-900 energy band to the $\mathrm{T_\mathrm{90}}$ \GBM{} durations. More than 60\% of GRBS recover more than 50\% of the $\mathrm{T_\mathrm{90}}$ in 50-300 keV for 34 GRBs found in 50-900 keV;20\% have a trigger duration 10\% longer than $\mathrm{T_\mathrm{90}}$. For 38 GRBs found in 4-900 keV, more than 80\% of GRBS recover more than 50\% of the $\mathrm{T_\mathrm{90}}$ in 50-300 keV and 30\% have a trigger duration 10\% longer than $\mathrm{T_\mathrm{90}}$, showing the contribution to longer soft emission. 


In conclusion, \FWBSB{} offers a unique and independent way to give a first classification  of the gamma-ray triggers based on their trigger duration. Even if the method is focused on the discovery of new triggers, it provides a duration of the prompt emission of the gamma-ray transients in several bands. Moreover, it helps to understand the spectral dynamics of the prompt emission that is much more complex with a smooth transition from the hard to soft X-ray bands with the different delays and durations across the multi-band triggers (mostly below  $\Delta_{\tilde{\tau_i},\tilde{\tau_j}}$=50 s). Further investigations are necessary to cleary identify the nature of the gamma-ray transients (e.g extra-galactic vs galactic sources) from the complete sample: we need an estimation of the location of the event, spectral properties and possible multi-wavelength counterpart.

\subsection{Detection in coincidence with others surveys}

\FWBSB{} also detects other events, especially in the softer band (4-50 keV). We perform some investigations by comparing \FWBSB{} events with other transients from gamma-ray surveys such as \href{https://gcn.gsfc.nasa.gov/gcn/fermi_gbm_subthresh_archive.html}{Fermi-GBM, subthresholds}, \href{https://swift.gsfc.nasa.gov/archive/grb_table/}{Swift-BAT}, \href{https://gcn.gsfc.nasa.gov/konus_grbs.html#tc1}{Konus-Wind}, \href{http://www.hxmt.org/images/GRB}{HXMT}, \href{http://www.hxmt.org/images/GRB}{CALET} and \href{https://gcn.gsfc.nasa.gov/maxi_grbs.html}{MAXI}. We allow +/- 30 second latency in timing between the two gamma-ray events. Our \FWBSB{} detects jointly with \GBM{} onboard analaysis 6 GRBs with Swift-BAT, 16+1 GRBs with Konus-Wind, 2 GRBs with MAXI, 2 GRBs with CALET and 6+1(galactic flare) GRBs with HXMT (see Table~\ref{tab:calGBMWBS}). Over the blind search period, \FWBSB{} does not detect gamma-ray events in coincidence with the 9 Swift-BAT GRBs, 22 X-ray MAXI flares, 6 HXMT GRBs,  7 Konus-Wind GRBs and no GRB170825 Agile GRB.
\smallbreak
We also detect 6 events among the 54 events in timing coincidence with the sub-threshold analysis of Fermi-\GBM{} listed in Table~\ref{FWBSBsubthreshold}. Taking $\tau_\mathrm{GBM}=54/46.5=1.16/\mathrm{day}$ and $\tau_\mathrm{WBS}=32/\mathrm{day}$ and within a window of $\Delta_\mathrm{coinc}=1 \mathrm{min}=6.9\times 10 ^{-4} \, \mathrm{day}$, the accidental coincidence rate is $R_\mathrm{fake,GBM}=0.03 \, \mathrm{/day}$ ($\approx$ 1.4 events for this analysis). Note that we consider \FWBSB{} and Fermi sub-threshold analysis as two independent studies, since the two approaches are very distinct, with an optimization on a short GRB investigation for the Fermi sub-threshold and a larger diversity GRB target for \FWBSB{}.

\begin{table*}
\caption{Results of the \FWBSB{} data analysis compared with
\href{https://gcn.gsfc.nasa.gov/gcn/fermi_gbm_subthresh_archive.html}{Fermi-GBM subthresholds}. From the 54 sub-thresholds triggers, 6  were also found by \FWBSB{} listed above. \label{FWBSBsubthreshold}}
\begin{tabular}{ccccccccc}
GBM Name &  Day & GBM Trigger  & \FWBSB{} Trigger & \FWBSB{} Duration  & \FWBSB{} Detector & \FWBSB{} SNR & \FWBSB{} Energy  & \FWBSB{} ref   \\
and ref & &  Time (UT) & Time (UT) & (s)  & involved &  & band (keV)  & \\
\hline
\href{https://gcn.gsfc.nasa.gov/gcn/notices_gbm_sub/556027113.fermi}{556027113} & 2018/08/15 & 11:58:28.82 & 11:59:10.66 & 8.064  & n2,na & 53.7 & 4-50, 4-900, 50-100 & \href{https://github.com/santier14/FWBSB/tree/master/August-September_2018/180815/GBMT556027113-WBS}{Link} \\
\href{https://gcn.gsfc.nasa.gov/gcn/notices_gbm_sub/525514019.fermi}{525514019} & 2017/08/27 & 08:06:54.63 & 08:06:01.28 & 0.512  & n5 & 9.8 & 4-50  & \href{https://github.com/santier14/FWBSB/tree/master/August_2017/170817/GBMT525514019-WBS}{Link} \\
\href{https://gcn.gsfc.nasa.gov/gcn/notices_gbm_sub/525127039.fermi}{525127039} & 2017/08/22 & 20:37:14.70 & 20:37:39.38 & 0.512  & n0 & 3.5 & 4-900 &
\href{https://github.com/santier14/FWBSB/tree/master/August_2017/170822/GBMT525514019-WBS}{Link} \\
\href{https://gcn.gsfc.nasa.gov/gcn/notices_gbm_sub/524798902.fermi}{524798902} & 2017/08/19 & 01:28:17.06 & 01:28:16.19 & 1.664  & n3 & 6.0 & 4-900 & \href{https://github.com/santier14/FWBSB/tree/master/August_2017/170819/GBMT525047873-WBS}{Link} \\
\href{https://gcn.gsfc.nasa.gov/gcn/notices_gbm_sub/524556701.fermi}{524556701} & 2017/08/16 & 06:11:36.23 & 06:11:54.18 & 2.048  & n7 & 6.9 & 4-900 & \href{https://github.com/santier14/FWBSB/tree/master/August_2017/170819/GBMT524556701-WBS}{Link} \\
\href{https://gcn.gsfc.nasa.gov/gcn/notices_gbm_sub/523661895.fermi}{523661895} & 2017/08/05 & 21:38:09.49 & 21:38:09.49 & 1.408  & n6 & 6.6 & 50-100, 50-900 & \href{https://github.com/santier14/FWBSB/tree/master/August_2017/170819/GBMT523661895-WBS}{Link} \\
\end{tabular}
\end{table*}

We detect four events correlated in time with Swift-BAT GRB but not found by onboard and offline GBM analysis.Table~\ref{FWBSBsubthreshold}. Taking $\tau_\mathrm{Swift}=90/365.25=0.25/\mathrm{day}$ and $\tau_\mathrm{WBS}=32/\mathrm{day}$ and within a window of $\Delta_\mathrm{coinc}=1 \mathrm{min}=6.9\times 10 ^{-4} \, \mathrm{day}$, the accidental coincidence rate is $R_\mathrm{fake,Swift}=0.006 \, \mathrm{/day}$ ($\approx$ 0.3 events for this analysis). 

\begin{table*}
\caption{Results of the \FWBSB{} data analysis compared with
\href{https://swift.gsfc.nasa.gov/archive/grb_table/}{Swift-BAT} when Fermi-GBM onboard did not trigger. From the 18+1(classified as galactic) Swift-BAT GRBs triggers, 6 have been detected jointly by Fermi-GBM and \FWBSB{} (see Table~\ref{tab:calGBMWBS}), and 4 with \FWBSB{} only reported here.}
\begin{tabular}{ccccccccc}
BAT Name &  Day & BAT Trigger  & \FWBSB{} Trigger & \FWBSB{} Duration  & \FWBSB{} Detector & \FWBSB{} SNR & \FWBSB{} Energy  & \FWBSB{} ref   \\
and ref & &  Time (UT) & Time (UT) & (s)  & involved &  & band (keV)  & \\
\hline
\href{https://gcn.gsfc.nasa.gov/other/766821.swift}{GRB170807A} & 2017/08/07 & 21:56:09.11 & 21:55:48.93 & 0.64  & n4,n5 & 31.2 & 4-50 & \href{https://github.com/santier14/FWBSB/tree/master/August_2017/170807/BATGRB170807A-WBS}{Link} \\
\href{https://gcn.gsfc.nasa.gov/other/851829.swift}{GRB180805A} & 2018/08/05 & 09:04:49.02 & 09:04:48.83 & 1.92  & n2 & 6.2 & 4-50 & \href{https://github.com/santier14/FWBSB/tree/master/August-September_2018/180805/BATGRB180805A-WBS}{Link} \\
\href{https://gcn.gsfc.nasa.gov/other/853824.swift}{GRB180818A} & 2018/08/18 & 03:12:04.03 & 03:12:03.01 & 6.66  & n0 & 6.1  & 4-50  & \href{https://github.com/santier14/FWBSB/tree/master/August-September_2018/180818/BATGRB180818A-WBS}{Link} \\
\href{https://gcn.gsfc.nasa.gov/other/859421.swift}{GRB180905A} & 2018/09/05 & 13:57:46.45 & 13:58:54.0 & 7.68  & n3,n8 & 7.0  & 4-900  & \href{https://github.com/santier14/FWBSB/tree/master/August-September_2018/180905/BATGRB180905A-WBS}{Link} \\
\end{tabular}
\end{table*}

We also detect four events with \FWBSB{} in timing coincidence with events found by MAXI below 20 keV.  Taking $\tau_\mathrm{MAXI}=28/46.5=0.60/\mathrm{day}$ and $\tau_\mathrm{FWBSB}=32/\mathrm{day}$ and within a window of $\Delta_\mathrm{coinc}=1 \mathrm{min}=6.9\times 10 ^{-4} \, \mathrm{day}$, the accidental coincidence rate is $R_\mathrm{fake,MAXI}=0.01 \, \mathrm{/day}$ ($\approx$ 0.5 events for this analysis). 

\smallbreak
In conclusion, either for Swift-BAT and MAXI triggers, \FWBSB{} independent searches offer the possibility to extend the spectral range of the X-ray transients and variable stars observed thanks to the broad energy band coverage of \GBM{}. For example, the joint analysis of GRB1808018A with BAT and GBM can help to estimate the $E_\mathrm{peak}$ parameter of the GRB spectrum. Also Swift localization is more accurate than GBM enabling spectroscopic follow-up and estimation of the redshift. In addition, comparison with multiple searches with independent false alarm rate estimation can confirm detections of weak signals. The joint \FWBSB{} and \href{https://gcn.gsfc.nasa.gov/gcn/fermi_gbm_subthresh_archive.html}{Fermi-GBM subthreshold analysis} are helpful for multi-messenger studies as for GW150914-GBM discussion \citep{2016ApJ...826L...6C}.

\subsection{Additional detection from \FWBSB{}}

\FWBSB{} also detects multiple events mainly in the soft band at a rate of 32 events per day in the 4-50 keV band. They are mainly due to variable stars and X-ray galactic flares. We investigate every event to check if it is not due to the approximate background estimation. However, without localization, it is hard to find the nature of the transient events. Nevertheless, we collect 171 events found by \FWBSB{} checking by hand the profile of those events. They are not found by other gamma-ray surveys compatible with GRB lightcurves provided in \href{https://github.com/santier14/FWBSB/commits/master}{here}. Among the 171 events, 95\% of the events are 4-50 keV events, 70\% of the events are in the 4-900 keV band and 0.5\% of the events are in the 50-900 and 50-100 keV bands. This sample has a low enough rate to be helpful in the future for cross-matching a sample of transients in other wavelengths such as orphan visible afterglows or other messengers such as gravitational waves and neutrinos.

\begin{table*}
\caption{Results of the \FWBSB{} data analysis compared with
\href{http://maxi.riken.jp/alert/novae/index.html}{MAXI} when Fermi-GBM onboard did not trigger. During the data analysis period, 28 events were detected by MAXI. 4 triggers were related to GRBs whereas the others are connected to flare stars. GRB170830A and GRB180809A were detected jointly by GBM-onboard and \FWBSB{} (see Table~\ref{tab:calGBMWBS})}.  
\begin{tabular}{cccccccccc}
MAXI Name  &  Day & MAXI Trig.  & \FWBSB{} Trig. & \FWBSB{} Durat.  & \FWBSB{} Detec. & \FWBSB{}  & \FWBSB{} Energy  & \FWBSB{}   \\
and ref  & &  Time (UT) & Time (UT) & (s)  & involved & SNR  & band (keV)  & ref \\
\hline
\href{http://maxi.riken.jp/alert/novae/8363174941/8363174941.html}{J1621-501}  & 2018/09/02 & 05:02:29 & 05:01:38.5 & 42.624  & n3,n6 & 6.8 & 4-900 & \href{https://github.com/santier14/FWBSB/tree/master/August-September_2018/180902/MAXI-180902}{Link} \\
\href{http://maxi.riken.jp/alert/novae/8335374685/8335374685.html}{J1621-501}  & 2018/08/05 & 11:37:03 & 11:37:19.42 & 0.512  & n5 & 33.3 & 4-50 & \href{https://github.com/santier14/FWBSB/tree/master/Calibration/180805/MAXI-170805}{Link} \\
\href{http://maxi.riken.jp/alert/novae/8332328054/8332328054.htm}{4U 1916-053}  & 2018/08/02 & 09:36:57 & 09:36:54.08 & 6.272  & na & 7.6 & 4-900 & \href{https://github.com/santier14/FWBSB/tree/master/Calibration/180802/MAXI-180802}{Link} \\
\href{http://maxi.riken.jp/alert/novae/7971325470/7971325470.htm}{H 1743-322} & 2017/08/06 & 09:54:14 & 09:54:27.26 & 23.296  & n5 & 9.7 & 4-900 & \href{https://github.com/santier14/FWBSB/tree/master/August_2017/170806/MAXI-20170806}{Link} \\
\hline
\end{tabular}
\end{table*}



\section{Discussion and Conclusion}

In this paper, we present \FWBSB{}, an automatic blind search for detecting of gamma-ray transients. It is particularly sensitive to spiky gamma-ray transients in the soft X-ray band. \FWBSB{} uses a unique technique of gamma-ray detection with wild binary segmentation of a timing series and less than intrinsic 10 parameters to tune.  We evaluated its performance using 60 days of Fermi-\GBM{} continuous daily data, for the 12 NaI detectors records after one week of calibration. In addition, multi-band searches encourage a low signal-to-noise identification of the spectral characteristics with hard peaks in the 50-900 keV band and shorter duration events, and longer duration events for soft band or detection of additional triggers generally delayed from the harder band. Independent continuous searches are a powerful tool to discard weak fluctuations of the background and weak gamma-ray transients. By increasing GBM's sensitivity, the detected GRB rate can be increased (as the count of localisation). In addition, this approach is helpful to validate low signal-to-noise GRB signals in GBM data in coincidence with gravitational-wave detections, which is important for multi-messenger studies of the violent Universe. The results performance of the \FWBSB{} analysis reveals the detection of 42 events in the 4-900, 50-100, 4-50 or 50-100 keV bands in coincidence in time with the on-board GBM analysis out of 44 gamma-ray transients found by GBM onboard. Short GRBs are less clearly identified as compared to long GRBs due to the minimal resolution of 128\,ms of \FWBSB{} compared to the 64 ms in on-board analysis. However, the multiple band peak detection in \FWBSB{} recovers more than 70\% of the duration of 7 out of 9 GRBs with $\mathrm{T90} \, > \, 100 s$ (50-300 keV) present in the sample. \FWBSB{} detects other gamma-ray transients at a rate of 30 events per day for the 4-50 keV band, 20 events per day for the 4-900 keV band, 1.2 event per day in the 50-100 keV and 50-900 keV bands. \FWBSB{} is able to recover GRBs found by the online on-board analysis and other surveys. But \FWBSB{} also finds events in time coincidence with other gamma-ray surveys while none are detected from online and offline \GBM{} searches:  one confident event seen by Swift-BAT (GRB180818A) and 4 others sub-threshold events found in coincident with Swift GRBs; there are also 4 \FWBSB{} events connected to MAXI galactic flares. A list of 171 events found by \FWBSB{} is unique and compatible with galactic or GRB light curve profiles. However, further investigations are required to determine the nature of the gamma-ray transient. For this purpose, new gamma-ray transient needs to be compatible with a extra-galactic source using rough localization \citep{2019ApJ...873...60B}, GRB spectrum (e.g Band model) and timing profile. This also helps for the deeper classification of the transient as  short and long GRBs or intermediate GRB classes \citep{1998ApJ...508..314M}, that might differ in terms of progenitors (coalescence of neutrons stars vs coalescence of massive stars).

Recently, a systematic analysis of \FWBSB{} over 24h is performed in case of a gravitational wave alert during the third observational campaign of LIGO-Virgo since April 2019. No gamma-ray trigger is found in the period of around the GW trigger time (-5s, 1h) for the first six months \citep{2019arXiv191011261A}. Indeed, this strategy gives a playground for multi-messenger searches with neutrino and gravitational-wave events as well as more broadly in time domain astronomy that will certainly help for identifying the nature of million of optical transients, including those from LSST \citep{2018MNRAS.481.4355D}.

\section*{Acknowledgements}
SA is supported by the CNES Postdoctoral Fellowship at Laboratoire AstroParticule et Cosmologie. SA and CL acknowledge the financial support of the UnivEarthS Labex program at Sorbonne Paris Cit\'e (ANR-10-LABX-0023 and ANR-11-IDEX-0005-02). We acknowledge the Laboratoire de l'Acc\'el\'erateur Lin\'eaire that provides founding KB for her internship. The work of PF was supported by the Engineering and Physical Sciences Research Council grant no.
EP/L014246/1.


\bsp	
\label{lastpage}

\bibliographystyle{mnras}
\bibliography{WBS}

\appendix

\section{}
\label{Appendixextable}

\begin{table*}
\caption{Calibration sample: results of the \FWBSB{} data analysis (August 2018) compared with the \href{https://gcn.gsfc.nasa.gov/fermi_grbs.html}{onboard GBM analysis} \citep{Bhat16}. The GRBs mentioned above were used as reference to calibrate \FWBSB{} parameters. \label{tab:calGBMWBS}}
\begin{tabular}{cccccccc}
Search & Detectors & \multicolumn{3}{c}{Trigger} & Energy  & GRB  & Ref  \\
& Involved & Time (UT) & Duration (s) & SNR &  band (keV)  & classification &  \\ 
\hline
\multicolumn{8}{c}{\textcolor{black}{bn180801276}} \\
\href{https://heasarc.gsfc.nasa.gov/FTP/fermi/data/gbm/bursts/2018/bn180801276/quicklook/glg_lc_all_bn180801276.gif}{GBM onboard} & n6,n7,n8,n9 & 06:37:04.51 &  0.064 & 5.7 & 47-291  & \multirow{2}{*}{long} & \href{https://gcn.gsfc.nasa.gov/other/554798228.fermi}{Notice}  \\
\FWBSB{}  & n6,n8 & 06:37:03.30  &  0.768 & 9.1 & 50-900  &  &  \href{https://github.com/santier14/FWBSB/tree/master/Calibration/180801/bn180801276-WBS}{Link} \\
\hline
\multicolumn{8}{c}{\textcolor{black}{bn180801492}} \\
\href{https://heasarc.gsfc.nasa.gov/FTP/fermi/data/gbm/bursts/2018/bn180801492/quicklook/glg_lc_all_bn180801492.gif}{GBM onboard} & n0,n1 & 11:48:42.17 & 1.024 & 4.8 & 47-291 &  \multirow{5}{*}{long} & \href{https://gcn.gsfc.nasa.gov/other/554816927.fermi}{Notice}  \\
\FWBSB{}  & n0,n3 & 11:48:41.30  &  8.064  & 17.2 & 4-50  & & \multirow{4}{*}{\href{https://github.com/santier14/FWBSB/tree/master/Calibration/180801/bn180801492-WBS}{Link}}  \\
\FWBSB{}  & n0, n3 & 11:48:43.80 &   5.12  & 9.8 & 50-100  & &   \\
\FWBSB{}  & n1 & 11:48:41.70  &  8.064 & 8.1 & 50-900  & &   \\
\FWBSB{}  & n0,n1,n3,n9 & 11:48:41.00  &  7.936  & 17.6 & 4-900  & &   \\
\hline
\multicolumn{8}{c}{\textcolor{black}{bn180803590}} \\
\href{https://heasarc.gsfc.nasa.gov/FTP/fermi/data/gbm/bursts/2018/bn180803590/quicklook/glg_lc_all_bn180803590.gif}{GBM onboard} & n3,n4 & 14:09:49.73 & 0.256 & 4.8 & 47-291 & \multirow{3}{*}{short} & \href{https://gcn.gsfc.nasa.gov/other/554998194.fermi}{Notice}  \\
\FWBSB{}  & n3 & 14:10:01.66  &  0.768  & 71.8 & 4-50  &  &  \multirow{2}{*}{\href{https://github.com/santier14/FWBSB/tree/master/Calibration/180803/bn180803590-WBS}{Link}} \\
\FWBSB{}  & n3 & 14:10:01.66  &  0.768  & 56.5 & 4-900  &  &   \\
\hline
\multicolumn{8}{c}{\textcolor{black}{bn180804554}} \\
\href{https://heasarc.gsfc.nasa.gov/FTP/fermi/data/gbm/bursts/2018/bn180804554/quicklook/glg_lc_all_bn180804554.gif}{GBM onboard} & n4,n5 & 13:17:41.44 & 2.048 & 4.9 & 47-291 & \multirow{5}{*}{long} & \href{https://gcn.gsfc.nasa.gov/other/555081466.fermi}{Notice}  \\
\FWBSB{} & n4,n5 & 13:17:48.99  &  17.664  & 16.5 & 4-50  & & \multirow{4}{*}{\href{https://github.com/santier14/FWBSB/tree/master/Calibration/180804/bn180804554-WBS}{Link}}  \\
\FWBSB{} & n1,n4,n5 & 13:17:45.80 &   5.12  & 17.4 & 50-100  & &   \\
\FWBSB{}  & n4,n5 & 13:17:38.24  &  29.696 & 32.2 & 50-900  & &   \\
\FWBSB{} & n0,n4,n5 & 13:17:37.60  &  7.936  & 24.0 & 4-900  & &   \\
\hline
\multicolumn{8}{c}{\textcolor{black}{bn180804765}} \\
\href{https://heasarc.gsfc.nasa.gov/FTP/fermi/data/gbm/bursts/2018/bn180804765/quicklook/glg_lc_all_bn180804765.gif}{GBM onboard} & n2,n5 & 18:22:19.65 & 4.096 & 5.2 & 47-291 & \multirow{3}{*}{long} & \href{https://gcn.gsfc.nasa.gov/other/555099744.fermi}{Notice}  \\
\FWBSB{} & n2,n5 & 18:22:13.18  &  17.92   & 9.0 & 4-50  & &  \multirow{2}{*}{\href{https://github.com/santier14/FWBSB/tree/master/Calibration/180804/bn180804765-WBS}{Link}} \\
\FWBSB{}  & n2,n5 & 18:22:14.34   &  15.744   & 10.6 & 4-900  & &   \\
\hline
\multicolumn{8}{c}{\textcolor{black}{bn180804909}} \\
\href{https://heasarc.gsfc.nasa.gov/FTP/fermi/data/gbm/bursts/2018/bn180804909/quicklook/glg_lc_all_bn180804909.gif}{GBM onboard} & n2,n5 & 21:49:02.18 & 4.096 & 4.7 & 47-291 & \multirow{3}{*}{long} & \href{https://gcn.gsfc.nasa.gov/other/555112147.fermi}{Notice}  \\
\FWBSB{}  & n1 & 21:48:04.67  &  36.224 & 8.2 & 50-900  & & \multirow{2}{*}{\href{https://github.com/santier14/FWBSB/tree/master/Calibration/180804/bn180804909-WBS}{Link}}  \\
\FWBSB{}  & n1,n2,n5 & 21:48:57.79  &  7.68  & 6.1 & 4-900  & &   \\
\hline
\multicolumn{8}{c}{\textcolor{black}{bn180804931}} \\
\href{https://heasarc.gsfc.nasa.gov/FTP/fermi/data/gbm/bursts/2018/bn180804931/quicklook/glg_lc_all_bn180804931.gif}{GBM onboard} & n8,nb & 22:20:38.28 & 0.256 & 4.9 & 47-291 & \multirow{5}{*}{long} & \href{https://gcn.gsfc.nasa.gov/other/554998194.fermi}{Notice}  \\
\FWBSB{}  & n8,nb & 22:20:37.10  &  13.056  & 10.2 & 4-50  & & \multirow{4}{*}{\href{https://github.com/santier14/FWBSB/tree/master/Calibration/180804/bn180804931-WBS}{Link}}  \\
\FWBSB{}  & n7,n8,nb & 22:20:37.41 &   15.9  & 12.0 & 50-100  & &   \\
\FWBSB{}  & n3,n4,n6 & 22:20:36.44  &  21.632 &  7.9 & 50-900  & &   \\
\FWBSB{}  & n3,n4,n5,n8,nb & 22:20:36.85  &  19.7  & 13.0 & 4-900  & &   \\
\hline
\multicolumn{8}{c}{\textcolor{black}{bn180805543}} \\
\href{https://heasarc.gsfc.nasa.gov/FTP/fermi/data/gbm/bursts/2018/bn180805543/quicklook/glg_lc_all_bn180805543.gif}{GBM onboard} & n6,n7 & 13:02:36.52 & 0.064 & 5.0 & 47-291 & \multirow{4}{*}{short} & \href{https://gcn.gsfc.nasa.gov/other/555166961.fermi}{Notice}  \\
\FWBSB{}  & n7 & 13:02:36.61   &  0.512 &  5.6 & 50-100  & & \multirow{3}{*}{\href{https://github.com/santier14/FWBSB/tree/master/Calibration/180805/bn180805543-WBS}{Link}}  \\
\FWBSB{}  & n6,n8,na & 13:02:36.61  &  0.512 &  4.9 & 50-900  & &   \\
\FWBSB{}  & n7 & 13:02:36.35   &  0.896  &  9.9 & 4-900  & &   \\
\href{https://gcn.gsfc.nasa.gov/notices_s/851855/BA/}{Swift-BAT} & - & 13:02:36.48 & 0.256 & 28.0 & 50-350 &   & \href{https://gcn.gsfc.nasa.gov/other/851855.swift}{Notice} \\
\hline
\multicolumn{8}{c}{bn180806665} \\
\href{https://heasarc.gsfc.nasa.gov/FTP/fermi/data/gbm/bursts/2018/bn180806665/quicklook/glg_lc_all_bn180806665.gif}{GBM onboard} & n6,n8 & 15:57:58.30 & 2.048 & 5.0 & 47-291 & \multirow{2}{*}{long} & \href{https://gcn.gsfc.nasa.gov/other/555263883.fermi}{Notice}  \\
\FWBSB{}  & \multicolumn{7}{c}{No detection}   \\
\hline
\multicolumn{8}{c}{\textcolor{black}{bn180806944}} \\
\href{https://heasarc.gsfc.nasa.gov/FTP/fermi/data/gbm/bursts/2018/bn180806944/quicklook/glg_lc_all_bn180806944.gif}{GBM onboard} & n7,n8 & 22:38:59.66 & 4.096 & 5.2 & 47-291 & \multirow{4}{*}{long} & \href{https://gcn.gsfc.nasa.gov/other/555099744.fermi}{Notice}  \\
\FWBSB{}  & n6,n7,n8,nb & 22:38:46.08  &  29.184 &  63-62 & 4-50  & & \multirow{3}{*}{\href{https://github.com/santier14/FWBSB/tree/master/Calibration/180806/bn180806944-WBS}{Link}}  \\
\FWBSB{}  & n6,n7,n8 & 22:38:57.35 &   18.304  & 41.8 & 50-100  & &   \\
\FWBSB{}  & n7,n8 & 22:38:57.47  &  24.32 &  62.9 & 50-900  & &   \\
\FWBSB{}  & n7,n8,n9,nb & 22:38:56.58 &  19.456  & 59.7 & 4-900  & &   \\
\hline
\multicolumn{8}{c}{\textcolor{black}{bn180807097}} \\
\href{https://heasarc.gsfc.nasa.gov/FTP/fermi/data/gbm/bursts/2018/bn180807097/quicklook/glg_lc_all_bn180807097.gif}{GBM onboard} & n8,na & 02:19:54.72 & 2.048 & 5.1 & 47-291 & \multirow{5}{*}{long} & \href{https://gcn.gsfc.nasa.gov/other/555301199.fermi}{Notice}  \\
\FWBSB{}  & na,nb & 02:19:41.51  &  38.656  & 8.9 & 4-50  &  &  \multirow{4}{*}{\href{https://github.com/santier14/FWBSB/tree/master/Calibration/180807/bn180807097-WBS}{Link}} \\
\FWBSB{}  & na & 02:19:52.77  &   4.992   & 5.8 & 50-100  & &   \\
\FWBSB{} & n8,na,nb & 02:19:51.36 &  4.096 &  6.3 & 50-900  & &   \\
\FWBSB{}  & n8,na,nb & 02:19:50.21  &  9.216  & 10.3 & 4-900  & &   \\
\end{tabular} 
\end{table*}

\begin{table*}
\caption{Results of the \FWBSB{} data analysis (August 2017, August 2018 and September 2018) compared with the \href{https://gcn.gsfc.nasa.gov/fermi_grbs.html}{onboard GBM analysis}. As an indication, other surveys that independently triggered on the same gamma-ray bursts as HXMT, Swift-BAT, Konus-Wind, Calet and MAXI are reported as well.\label{tabresGBMWBS2}}
\begin{tabular}{cccccccc}
Search & Detectors & \multicolumn{3}{c}{Trigger} & Energy  & GRB  & Ref  \\
& Involved & Time (UT) & Duration (s) & SNR (av) &  band (keV)  & classification &  \\
\multicolumn{8}{c}{\textcolor{black}{August 2017}} \\
\hline
\multicolumn{8}{c}{\textcolor{black}{bn170801690}} \\
\href{https://heasarc.gsfc.nasa.gov/FTP/fermi/data/gbm/bursts/2017/bn170801690/quicklook/glg_lc_all_bn170801690.gif}{GBM onboard} & n8,nb & 16:33:43.40 & 2.048 & 4.5 & 47-291 &  \multirow{2}{*}{long} & \href{https://gcn.gsfc.nasa.gov/other/523298028.fermi}{Notice}  \\
\FWBSB{} & n7,n8,nb & 16:33:40.80 &  4.224  & 7.4 & 4-900   & &  \href{https://github.com/santier14/FWBSB/tree/master/August_2017/170801/bn170801690-WBS}{Link}  \\
\hline
\multicolumn{8}{c}{\textcolor{black}{bn170802638}} \\
\href{https://heasarc.gsfc.nasa.gov/FTP/fermi/data/gbm/bursts/2017/bn170802638/quicklook/glg_lc_all_bn170802638.gif}{GBM onboard} & n6,n8 & 15:18:24.80 & 0.064 & 5.0 & 47-291 &  \multirow{7}{*}{short} &  \href{https://gcn.gsfc.nasa.gov/other/523379909.fermi}{Notice} \\
\FWBSB{} & n7 &   15:18:26.24  &   0.512   &   11.0    & 4-50 & &  \multirow{4}{*}{\href{https://github.com/santier14/FWBSB/tree/master/August_2017/170802/bn170802638-WBS}{Link}} \\
\FWBSB{} & n3,n4,n6,n7,n8,n9,na,nb & 15:18:26.37 & 1.152 & 16.4 & 4-900 & &   \\
\FWBSB{} & n3,n7 & 15:18:26.38  & 0.896  & 14.0 & 50-100 & &   \\
\FWBSB{} & n3,n4,n6,n8,n9,na,nb & 15:18:26.24 & 1.152 & 19.1 & 50-900 & &   \\
\href{http://www.hxmt.org/images/GRB/HEB170802637_lc.jpg}{HXMT} & - & 15:18:26.00 & 0.977 & - & 80-800 &   & -  \\
\href{http://www.ioffe.ru/LEA/kw/triggers/2017/kw20170802_55108.html}{Konus-WIND} & - & 15:18:28.10 & 0.38 & - & 50-200 &   & -  \\
\hline
\multicolumn{8}{c}{\textcolor{black}{bn170803172}} \\
\href{https://heasarc.gsfc.nasa.gov/FTP/fermi/data/gbm/bursts/2017/bn170803172/quicklook/glg_lc_all_bn170803172.gif}{GBM onboard} & n0,n1 & 04:07:15.75 & 2.048 & 5.4 & 47-291 &  \multirow{2}{*}{long} & \href{https://gcn.gsfc.nasa.gov/other/523426040.fermi}{Notice}  \\
\FWBSB{} & n1 &  04:07:14.63 & 1.408 & 6.9 & 50-100  & &  \multirow{1}{*}{\href{https://github.com/santier14/FWBSB/tree/master/August_2017/170803/bn170803172-WBS}{Link}} \\
\hline
\multicolumn{8}{c}{\textcolor{black}{bn170803415}} \\
\href{https://heasarc.gsfc.nasa.gov/FTP/fermi/data/gbm/bursts/2017/bn170803415/quicklook/glg_lc_all_bn170803415.gif}{GBM onboard} & n8,nb & 09:57:44.26 & 1.024 & 5.0 & 47-291 &  \multirow{5}{*}{long} & \href{https://gcn.gsfc.nasa.gov/other/523447069.fermi}{Notice}  \\
\FWBSB{} & n7,n8,nb &  09:57:40.80 & 77.056 & 12.7 &  4-50 & & \multirow{4}{*}{\href{https://github.com/santier14/FWBSB/tree/master/August_2017/170803/bn170803415-WBS}{Link}} \\
\FWBSB{} & n7,n8,nb & 09:57:41.82 & 62.848 & 15.0 & 4-900 & &   \\
\FWBSB{} & n6,n8,nb & 09:57:42.60 & 50.297 & 10.0 &  50-100 & &   \\
\FWBSB{} & n6,n7,n8,nb & 09:57:40.81 & 72.704 & 7.3  & 50-900 & &   \\
\hline
\multicolumn{8}{c}{\textcolor{black}{bn170803729}} \\
\href{https://heasarc.gsfc.nasa.gov/FTP/fermi/data/gbm/bursts/2017/bn170803729/quicklook/glg_lc_all_bn170803729.gif}{GBM onboard} & n1,n3 & 17:30:27.11 & 0.064 & 5.6 & 47-291 &  \multirow{5}{*}{long} & \href{https://gcn.gsfc.nasa.gov/other/523474232.fermi}{Notice}  \\
\FWBSB{} & n0,n1,n3,n4,n5 &  17:30:26.75 & 4.480   &   14.6  & 4-50 & & \multirow{4}{*}{\href{https://github.com/santier14/FWBSB/tree/master/August_2017/170803/bn170803729-WBS}{Link}}  \\
\FWBSB{} & n0,n1,n3,n4,n5,n6 & 17:30:26.62  & 4.992  & 19.5 & 4-900 & &   \\
\FWBSB{} & n0,n1,n2,n3,n4,n5,n6 & 17:30:26.63 & 2.944 & 14.9 & 50-100 & &   \\
\FWBSB{} & n0,n1,n3,n5,n6 & 17:30:26.63  & 4.736 & 42.5  & 50-900 & &   \\
\href{https://gcn.gsfc.nasa.gov/other/766081.swift}{Swif-BAT} & - & 17:30:27.17 & 0.064  & 32.6 & 25-100 &   & \href{https://gcn.gsfc.nasa.gov/notices_s/766081/BA/}{Notice}  \\
\href{http://www.ioffe.ru/LEA/kw/triggers/2017/kw20170803_63022.html}{Konus-WIND} & - & 17:30:22.76 & 2.054 & - & 50-200 &   & -  \\
\multicolumn{8}{c}{\textcolor{black}{bn170804911}} \\
\href{https://heasarc.gsfc.nasa.gov/FTP/fermi/data/gbm/bursts/2017/bn170804911/quicklook/glg_lc_all_bn170804911.gif}{GBM onboard} & n4,n8 & 21:52:21.46 & 2.048 & 4.7 & 47-291 &  \multirow{5}{*}{long} & \href{https://gcn.gsfc.nasa.gov/other/523576346.fermi}{Notice}  \\
\FWBSB{} & n4,n8 &  21:52:09.15  &  26.368 & 14.5 & 4-50 & & \multirow{4}{*}{\href{https://github.com/santier14/FWBSB/tree/master/August_2017/170804/bn170804911-WBS}{Link}}  \\
\FWBSB{} & n4,n8 & 21:52:17.98   & 14.720 & 18.8 & 4-900 & &   \\
\FWBSB{} & n4,n8 & 21:52:20.16  & 11.520 & 10.9  & 50-100 & &   \\
\FWBSB{} & n4,n8 & 21:52:20.80 & 10.368 & 11.3 & 50-900 & &   \\
\hline
\multicolumn{8}{c}{\textcolor{black}{bn170805901}} \\
\href{https://heasarc.gsfc.nasa.gov/FTP/fermi/data/gbm/bursts/2017/bn170805901/quicklook/glg_lc_all_bn170805901.gif}{GBM onboard} & n6,n9 & 21:37:49.59 & 1.024 & 4.5 & 47-291 & \multirow{3}{*}{long} & \href{https://gcn.gsfc.nasa.gov/other/523661874.fermi}{Notice}  \\
\FWBSB{} & n6,n7,n9 & 21:37:48.22 & 3.072  & 7.4 & 4-900 & &  \multirow{2}{*}{\href{https://github.com/santier14/FWBSB/tree/master/August_2017/170805/bn170805901-WBS}{Link}}  \\
\FWBSB{} & n6,n9 & 21:37:48.86 & 21.76 & 3.8  & 50-900 & &   \\
\multicolumn{8}{c}{\textcolor{black}{bn170808065}} \\
\href{https://heasarc.gsfc.nasa.gov/FTP/fermi/data/gbm/bursts/2017/bn170808065/quicklook/glg_lc_all_bn170808065.gif}{GBM onboard} & n2,na & 01:34:09.39 & 0.512 & 4.6 & 47-291 &  \multirow{5}{*}{long} & \href{https://gcn.gsfc.nasa.gov/other/523661874.fermi}{Notice}  \\
\FWBSB{} & n2,na & 01:34:09.85 & 7.552 & 11.3 & 4-50 & & \multirow{4}{*}{\href{https://github.com/santier14/FWBSB/tree/master/August_2017/170808/bn170808065-WBS}{Link}}  \\
\FWBSB{} & n1,n2,n8,n9,na & 01:34:07.67 & 5.760 & 13.3  & 4-900 & &   \\
\FWBSB{} &  n1,na & 01:34:08.37  & 5.120  & 12.1  & 50-100 & &   \\
\FWBSB{} & n1,n2,n7,n8,n9,na & 01:34:07.22  & 6.016 & 11.1 & 50-900 & &   \\
\href{http://www.ioffe.ru/LEA/kw/triggers/2017/kw20170808_05653.html}{Konus-WIND} & - & 01:34:13.07 & 5.25 & - & 50-200 &   & -  \\
\hline
\multicolumn{8}{c}{\textcolor{black}{bn170808936}} \\
\href{https://heasarc.gsfc.nasa.gov/FTP/fermi/data/gbm/bursts/2017/bn170808936/quicklook/glg_lc_all_bn170808936.gif}{GBM onboard} & n1,n5 & 22:27:43.10 & 0.512 & 5.2 & 47-291 &  \multirow{5}{*}{long} & \href{https://gcn.gsfc.nasa.gov/other/523924068.fermi}{Notice}  \\
\FWBSB{} & n9,na & 22:27:01.18 & 43.392  & 6.3 & 4-50  & & \multirow{4}{*}{\href{https://github.com/santier14/FWBSB/tree/master/August_2017/170808/bn170808936-WBS}{Link}}   \\
\FWBSB{} & n9,na & 22:27:02.85 & 40.704   & 3.9  & 4-900 & &   \\
\FWBSB{} & n0,n1,n2,n3,n5& 22:28:06.73 & 3.58 & 10.2 & 50-100 & &   \\
\FWBSB{} & n1,n5 & 22:28:05.83 & 45.568 & 1.0      & 50-900  & &   \\
\href{http://www.ioffe.ru/LEA/kw/triggers/2017/kw20170808_80859.html}{Konus-WIND} & - & 22:27:39.19 & 17.718 & - & 50-200 &   & \href{https://gcn.gsfc.nasa.gov/other/523661874.fermi}{Notice} \\
\end{tabular} 
\end{table*}

\begin{table*}
\contcaption{Results of the \FWBSB{} data analysis (August 2017, August 2018 and September 2018) compared with the \href{https://gcn.gsfc.nasa.gov/fermi_grbs.html}{onboard GBM analysis}. As an indication, other surveys that independently triggered on the same gamma-ray bursts as HXMT, Swift-BAT, Konus-Wind, Calet and MAXI are reported as well.}
\begin{tabular}{cccccccc}
Search & Detectors & \multicolumn{3}{c}{Trigger} & Energy  & GRB  & Ref  \\
& Involved & Time (UT) & Duration (s) & SNR (av) &  band (keV)  & classification &  \\
\multicolumn{8}{c}{\textcolor{black}{August 2017}} \\
\hline
\multicolumn{8}{c}{\textcolor{black}{bn170810918}} \\
\href{https://heasarc.gsfc.nasa.gov/FTP/fermi/data/gbm/bursts/2017/bn170810918/quicklook/glg_lc_all_bn170810918.gif}{GBM onboard} & n0,n1 & 22:01:41.58 & 0.256 & 4.5 & 47-291 &  \multirow{4}{*}{long} & \href{https://gcn.gsfc.nasa.gov/other/524095306.fermi}{Notice}  \\
\FWBSB{} & n0,n1 & 22:01:16.10  & 70.4 &  9.2  & 4-900 & & \multirow{3}{*}{\href{https://github.com/santier14/FWBSB/tree/master/August_2017/170810/bn170810918-WBS}{Link}}   \\   
\FWBSB{} & n0,n1,n2 & 22:01:34.41 & 47.232  & 7.9 & 50-100 & &   \\ 
\FWBSB{} & n0,n1,n2 & 22:01:31.20 & 54.016 & 11.9 & 50-900 & &   \\ 
\href{https://gcn.gsfc.nasa.gov/other/767284.swift}{Swif-BAT} & - & 22:01:41.06 & 1.024  & 23.32 & 25-100 &   & \href{https://gcn.gsfc.nasa.gov/notices_s/767284/BA/}{Notice}  \\
\hline
\multicolumn{8}{c}{\textcolor{black}{bn170813051}} \\
\href{https://heasarc.gsfc.nasa.gov/FTP/fermi/data/gbm/bursts/2017/bn170813051/quicklook/glg_lc_all_bn170813051.gif}{GBM onboard} & n0,n1,n2,n3 & 01:13:08.80 & 0.512 & 4.6 & 47-291 & \multirow{4}{*}{long} & \href{https://gcn.gsfc.nasa.gov/other/524279593.fermi}{Notice}  \\
\FWBSB{} & n1,n2 & 01:13:06.11 & 23.168 & 6.4   & 4-50 & & \multirow{3}{*}{\href{https://github.com/santier14/FWBSB/tree/master/August_2017/170813/bn170813051-WBS}{Link}}  \\    
\FWBSB{} & n0,n1,n2,n4 & 01:13:06.62 & 25.856 & 11.7   & 4-900 & &   \\     
\FWBSB{} & n0,n1 & 01:13:06.63 & 30.208 & 14.9      & 50-900 & &   \\    
\href{https://gcn.gsfc.nasa.gov/other/767563.swift}{Swif-BAT} & - & 01:13:16.48 & 64  & 12.3 & 15-50 &   & \href{https://gcn.gsfc.nasa.gov/notices_s/767563/BA/}{Notice}  \\
\hline
\multicolumn{8}{c}{\textcolor{black}{bn170816258}} \\
\href{https://heasarc.gsfc.nasa.gov/FTP/fermi/data/gbm/bursts/2017/bn170816258/quicklook/glg_lc_all_bn170816258.gif}{GBM onboard} & n5,n8 & 06:11:11.88 & 2.048 & 4.7 & 47-291 &  long & \href{https://gcn.gsfc.nasa.gov/other/524556676.fermi}{Notice}  \\
\FWBSB{} & \multicolumn{7}{c}{No detection} \\
\hline
\multicolumn{8}{c}{\textcolor{black}{bn170816599}} \\
\href{https://heasarc.gsfc.nasa.gov/FTP/fermi/data/gbm/bursts/2017/bn170816599/quicklook/glg_lc_all_bn170816599.gif}{GBM onboard} & n7,n8,na,nb & 14:23:03.96 & 0.016 & 10.6 & 47-291 & \multirow{4}{*}{short} & \href{https://gcn.gsfc.nasa.gov/other/524586188.fermi}{Notice}  \\
\FWBSB{} & n6,n7,n8,na,nb & 14:23:03.74  & 2.176  &   13.9 & 4-900 & &  \multirow{2}{*}{\href{https://github.com/santier14/FWBSB/tree/master/August_2017/170816/bn170816599-WBS}{Link}}  \\       
\FWBSB{} & n3,n6,n7,n8,n9,na,nb &  14:23:03.62 & 0.896 & 17.1    & 50-900 & &   \\
\href{http://cgbm.calet.jp/cgbm_trigger/flight/1186927993/}{CALET} & - & 14:23:03.81 & 0.5 & 9.8 & 40-230 &   & \href{https://gcn.gsfc.nasa.gov/gcn3/21615.gcn3}{Notice}  \\
\href{http://www.ioffe.ru/LEA/kw/triggers/2017/kw20170816_51789.html}{Konus-Wind} & - & 14:23:09.11 & 1.486 & - & 50-200 &   & \href{http://www.ioffe.ru/LEA/GRBs/GRB170816_T51789/}{Notice}  \\
\hline
\multicolumn{8}{c}{\textcolor{black}{bn170817529}} \\
\href{https://heasarc.gsfc.nasa.gov/FTP/fermi/data/gbm/bursts/2017/bn170817529/quicklook/glg_lc_all_bn170817529.gif}{GBM onboard} & n1,n2,n5 & 12:41:06.47 & 0.256 & 4.8 & 47-291 & short & \href{https://gcn.gsfc.nasa.gov/other/524666471.fermi}{Notice}  \\
\FWBSB{} & \multicolumn{7}{c}{No detection} \\
\hline
\multicolumn{8}{c}{\textcolor{black}{bn170817908}} \\
\href{https://heasarc.gsfc.nasa.gov/FTP/fermi/data/gbm/bursts/2017/bn170817908/quicklook/glg_lc_all_bn170817908.gif}{GBM onboard} & n2,n5 & 21:47:34.43 & 0.032 & 7.9 & 47-291 & \multirow{5}{*}{long} & \href{https://gcn.gsfc.nasa.gov/other/524699259.fermi}{Notice}  \\
\FWBSB{} & n0,n1,n2,n5 & 21:47:33.31 &   3.456  & 11.4 & 4-50 & & \multirow{4}{*}{\href{https://github.com/santier14/FWBSB/tree/master/August_2017/170817/bn170817908-WBS}{Link}}    \\   
\FWBSB{} & n0,n1,n2,n5,n9 & 21:47:33.31  & 0.256 & 11.5 & 4-900 & &   \\ 
\FWBSB{} & n0,n1,n2,n4,n5,n8,n9 & 21:47:32.98  & 3.584 & 11.4 & 50-100 & &   \\ 
\FWBSB{} & n0,n1,n2,n5 & 21:47:33.22 & 0.256  & 10.6 & 50-900  & &   \\ \href{http://www.hxmt.org/images/GRB/HEB170817908_lc.jpg}{HXMT} & - & 21:47:34.00 & 2.67 & - & 80-800 &   & \href{https://gcn.gsfc.nasa.gov/gcn3/21593.gcn3}{Notice}  \\
\href{http://www.ioffe.ru/LEA/kw/triggers/2017/kw20170817_78454.html}{Konus-WIND} & - & 21:47:34.02 & 2.694 & - & 50-200 &   & -  \\
\hline
\multicolumn{8}{c}{\textcolor{black}{bn170818137}} \\
\href{https://heasarc.gsfc.nasa.gov/FTP/fermi/data/gbm/bursts/2017/bn170818137/quicklook/glg_lc_all_bn170818137.gif}{GBM onboard} & n8,nb & 03:17:19.98 & 0.064 & 5.8 & 47-291 & short & \href{https://gcn.gsfc.nasa.gov/other/524719044.fermi}{Notice}  \\
\FWBSB{}  &  n8 & 03:17:19.68  & 0.896 & 6.0  & 50-100  & & \multirow{1}{*}{\href{https://github.com/santier14/FWBSB/tree/master/August_2017/170818/bn170818137-WBS}{Link}}  \\    
\hline
\multicolumn{8}{c}{\textcolor{black}{bn170821265}} \\
\href{https://heasarc.gsfc.nasa.gov/FTP/fermi/data/gbm/bursts/2017/bn170821265/quicklook/glg_lc_all_bn170821265.gif}{GBM onboard} & n1,n9 & 06:22:00.85 & 2.048 & 4.9 & 47-291 & \multirow{5}{*}{long} & \href{https://gcn.gsfc.nasa.gov/other/524989325.fermi}{Notice}  \\
\FWBSB{} & n0,n1,n9,na &  06:22:02.18 & 21.632 & 8.1  & 4-50 & & \multirow{4}{*}{\href{https://github.com/santier14/FWBSB/tree/master/August_2017/170821/bn170821265-WBS}{Link}}   \\
\FWBSB{} & n1,n9 & 06:21:58.47 & 25.344  & 10.9  & 4-900  & &   \\
\FWBSB{} & n1,n9 & 06:21:58.34 & 10.368  & 8.5 & 50-100 & &   \\
\FWBSB{} & n9 & 06:21:57.70 & 11.008  & 8.0 & 50-900 & &   \\
\hline
\multicolumn{8}{c}{\textcolor{black}{bn170825307}} \\
\href{https://heasarc.gsfc.nasa.gov/FTP/fermi/data/gbm/bursts/2017/bn170825307/quicklook/glg_lc_all_bn170825307.gif}{GBM onboard} & n3,n7 & 07:22:01.42 & 1.024 & 4.6 & 47-291 & \multirow{5}{*}{long} & \href{https://gcn.gsfc.nasa.gov/other/525338526.fermi}{Notice}  \\
\FWBSB{} & n3,n6,n8 & 07:22:00.26 & 8.448 & 8.7   & 4-50 & & \multirow{4}{*}{\href{https://github.com/santier14/FWBSB/tree/master/August_2017/170825/bn170825307-WBS}{Link}}   \\
\FWBSB{} & n3,n4,n6,n7,n8 & 07:22:00.77 & 5.76 & 16.7 & 4-900 & &   \\
\FWBSB{} & n3,n4,n6,n7 & 07:22:00.52  & 6.016  & 9.8 &  50-100 & &   \\
\FWBSB{} & n3,n4,n5,n6,n7,n8,na & 07:21:59.62  &  6.784  & 13.5 & 50-900 & &   \\
\href{http://www.hxmt.org/images/GRB/HEV170825306_lc.jpg}{HXMT} & - & 07:22:03.00 & 4.77 & - & 80-800 &   & -  \\
\href{http://www.ioffe.ru/LEA/kw/triggers/2017/kw20170825_26526.html}{Konus-WIND} & - & 07:22:06.22 & 6.345 & - & 50-200 &   & -  \\
\hline
\multicolumn{8}{c}{\textcolor{black}{bn170825500}} \\
\href{https://heasarc.gsfc.nasa.gov/FTP/fermi/data/gbm/bursts/2017/bn170825500/quicklook/glg_lc_all_bn170825500.gif}{GBM onboard} & n6,n8 & 12:00:06.00 & 0.512 & 5.9 & 47-291 & \multirow{5}{*}{long} & \href{https://gcn.gsfc.nasa.gov/other/525355210.fermi}{Notice}  \\
\FWBSB{} & n4,n5,n6,n7,n8,nb & 12:00:05.18  & 9.344 & 44.0  & 4-50 & & \multirow{4}{*}{\href{https://github.com/santier14/FWBSB/tree/master/August_2017/170825/bn170825500-WBS}{Link}}  \\
\FWBSB{} & n3,n4,n5,n6,n9,nb &  12:00:05.31 & 3.712 & 13.9  & 4-900 & &   \\
\FWBSB{} & n6,n7,nb & 11:59:31.26 &  42.624 & 6.8 & 50-100 & &   \\
\FWBSB{} & n3,n6,n7,n8,n9,nb & 12:00:01.60 & 7.552 & 9.4 & 50-900 & &   \\
\href{http://www.ioffe.ru/LEA/kw/triggers/2017/kw20170825_43214.html}{Konus-WIND} & - & 12:00:14.04 & 6.791 & - & 50-200 &   & -  \\
\end{tabular}
\end{table*}

\begin{table*}
\contcaption{Results of the \FWBSB{} data analysis (August 2017, August 2018 and September 2018) compared with the \href{https://gcn.gsfc.nasa.gov/fermi_grbs.html}{onboard GBM analysis}. As an indication, other surveys that independently triggered on the same gamma-ray bursts as HXMT, Swift-BAT, Konus-Wind, Calet and MAXI are reported as well.}
\begin{tabular}{cccccccc}
Search & Detectors & \multicolumn{3}{c}{Trigger} & Energy  & GRB  & Ref  \\
& Involved & Time (UT) & Duration (s) & SNR (av) &  band (keV)  & classification &  \\
\multicolumn{8}{c}{\textcolor{black}{August 2017}} \\
\hline
\multicolumn{8}{c}{\textcolor{black}{bn170825784}} \\
\href{https://heasarc.gsfc.nasa.gov/FTP/fermi/data/gbm/bursts/2017/bn170825784/quicklook/glg_lc_all_bn170825784.gif}{GBM onboard} & n3,n5 & 18:49:11.04 & 2.048 & 5.1 & 47-291 & \multirow{5}{*}{long} & \href{https://gcn.gsfc.nasa.gov/other/525379756.fermi}{Notice}  \\
\FWBSB{} & n3,n4 & 18:48:46.06 &    39.552 & 19.0   & 4-50  & & \multirow{4}{*}{\href{https://github.com/santier14/FWBSB/tree/master/August_2017/170825/bn170825784-WBS}{Link}}  \\ 
\FWBSB{} & n4,n6,n7 & 18:48:48.11 & 38.656   & 15.8 & 4-900 & &   \\
\FWBSB{} & n3,n4,n6 & 18:48:46.03 & 36.48 & 15.0 & 50-100 & &   \\
\FWBSB{} & n3,n4,n6 & 18:49:03.74 & 18.944 & 15.6 & 50-900 & &   \\
\hline
\multicolumn{8}{c}{\textcolor{black}{bn170826369}} \\
\href{https://heasarc.gsfc.nasa.gov/FTP/fermi/data/gbm/bursts/2017/bn170826369/quicklook/glg_lc_all_bn170826369.gif}{GBM onboard} & n1,n2 & 08:51:07.51 & 0.016 & 7.5 & 47-291 & short & \href{https://gcn.gsfc.nasa.gov/other/525430272.fermi}{Notice}  \\
\FWBSB{} & \multicolumn{7}{c}{No data available} \\
\href{http://www.ioffe.ru/LEA/kw/triggers/2017/kw20170826_31868.html}{Konus-WIND} & - & 08:51:08.41 & 0.114 & - & 50-200 &   & \href{https://gcn.gsfc.nasa.gov/gcn3/21774.gcn3}{Notice}  \\
\hline
\multicolumn{8}{c}{\textcolor{black}{bn170826819}} \\
\href{https://heasarc.gsfc.nasa.gov/FTP/fermi/data/gbm/bursts/2017/bn170826819/quicklook/glg_lc_all_bn170826819.gif}{GBM onboard} & na,nb & 19:38:56.48 & 1.024 & 4.5 & 47-291 & long & \href{https://gcn.gsfc.nasa.gov/other/525469141.fermi}{Notice}  \\
\href{http://www.hxmt.org/images/GRB/HEB170826818_lc.jpg}{HXMT} & - & 19:38:58.00 & 10.12 & - & 80-800 &   & \href{https://gcn.gsfc.nasa.gov/gcn3/21739.gcn3}{Notice}  \\
\FWBSB{} & \multicolumn{7}{c}{No data available} \\
\href{http://www.ioffe.ru/LEA/kw/triggers/2017/kw20170826_70742.html}{Konus-WIND} & - & 19:39:02.55 & 9.758 & - & 50-200 &   & \href{https://gcn.gsfc.nasa.gov/gcn3/21767.gcn3}{Notice}  \\
\hline
\multicolumn{8}{c}{\textcolor{black}{bn170827818}} \\
\href{https://heasarc.gsfc.nasa.gov/FTP/fermi/data/gbm/bursts/2017/bn170827818/quicklook/glg_lc_all_bn170827818.gif}{GBM onboard} & n0,n2 & 19:38:04.46 & 0.512 & 4.8 & 47-291 & short & \href{https://gcn.gsfc.nasa.gov/other/525555489.fermi}{Notice}  \\
\FWBSB{} & n0,n1 &   19:38:03.44     & 1.408  & 9.9-10.0  & 4-900 & & \multirow{2}{*}{\href{https://github.com/santier14/FWBSB/tree/master/August_2017/170827/bn170827818-WBS}{Link}}  \\  
\FWBSB{} & n0,n2,n5 & 19:38:04.10  & 0.768 & 6.1  & 50-900 & &   \\ 
\href{http://www.ioffe.ru/LEA/kw/triggers/2017/kw20170827_70686.html}{Konus-WIND} & - & 19:38:06.84 & 0.178 & - & 50-200 &   & \href{https://gcn.gsfc.nasa.gov/gcn3/21775.gcn3}{Notice}  \\
\hline
\multicolumn{8}{c}{\textcolor{black}{bn170829414}} \\
\href{https://heasarc.gsfc.nasa.gov/FTP/fermi/data/gbm/bursts/2017/bn170829414/quicklook/glg_lc_all_bn170829414.gif}{GBM onboard} & n7,nb & 09:56:30.58 & 0.512 & 4.8 & 47-291 & \multirow{7}{*}{long} & \href{https://gcn.gsfc.nasa.gov/other/525693395.fermi}{Notice}  \\
\FWBSB{} & n6,n7,n9 &  09:56:30.15 & 47.616 & 13.2  & 4-50 & & \multirow{6}{*}{\href{https://github.com/santier14/FWBSB/tree/master/August_2017/170829/bn170829414-WBS}{Link}}  \\   
\FWBSB{} & n6,n7 & 09:58:04.99 & 3.968 & 7.0  & 4-50 & &   \\  
\FWBSB{} & n6,n7,n8,n9,nb & 09:56:29.25 & 49.28  & 16.7 & 4-900 & &   \\     
\FWBSB{} & n6,n7 & 09:58:05.12 & 3.84  & 6.3 & 4-900 & &   \\
\FWBSB{} &  n6,n7,n8,n9 & 09:56:29.00  & 43.776 & 13.1 & 50-100 & &   \\
\FWBSB{} & n6,n7,n8,n9,nb & 09:56:29.25  &    43.648 & 15.4 & 50-900 & &   \\
\hline
\multicolumn{8}{c}{\textcolor{black}{bn170829674}} \\
\href{https://heasarc.gsfc.nasa.gov/FTP/fermi/data/gbm/bursts/2017/bn170829674/quicklook/glg_lc_all_bn170829674.gif}{GBM onboard} & n8,n9,nb & 16:10:03.16 & 0.256 & 4.9 & 47-291 & \multirow{5}{*}{long} & \href{https://gcn.gsfc.nasa.gov/other/525715808.fermi}{Notice}  \\
\FWBSB{} & n6,n7,n9 & 16:10:02.56 & 43.776 & 9.3 & 4-50 & & \multirow{4}{*}{\href{https://github.com/santier14/FWBSB/tree/master/August_2017/170829/bn170829674-WBS}{Link}}   \\
\FWBSB{} & n6,n7,n8,n9,na,nb & 16:10:02.82 & 45.696 & 11.9 & 4-900 & &   \\
\FWBSB{} & n6,n7,n9,na,nb & 16:10:02.82  & 10.624 & 11.9  & 50-100 & &   \\
\FWBSB{} & n6,n7,n8,n9,na,nb & 16:10:02.82 & 13.312  & 13.2 &  50-900 & &   \\
\href{http://www.ioffe.ru/LEA/kw/triggers/2017/kw20170829_58207.html}{Konus-WIND} & - & 16:10:07.68 & 44.744 & - & 50-200 &   & -  \\
\hline
\multicolumn{8}{c}{\textcolor{black}{bn170830069}} \\
\href{https://heasarc.gsfc.nasa.gov/FTP/fermi/data/gbm/bursts/2017/bn170830069/quicklook/glg_lc_all_bn170830069.gif}{GBM onboard} & n7,nb & 01:38:44.23 & 1.024 & 4.7 & 47-291 & \multirow{5}{*}{long} & \href{https://gcn.gsfc.nasa.gov/other/525749929.fermi}{Notice}  \\
\FWBSB{} & n6,n9 & 01:39:03.62  & 1.664 & 6.8 &  4-50  & & \multirow{4}{*}{\href{https://github.com/santier14/FWBSB/tree/master/August_2017/170830/bn170830069-WBS}{Link}}  \\
\FWBSB{} & n4,n6,n8,n9 & 01:38:59.65    & 14.848  & 11.7 & 4-900 & &   \\   
\FWBSB{} & n9,nb & 01:39:00.06 & 13.440 & 10.7 & 50-100 & &   \\ 
\FWBSB{} & n6,n7,n8,n9,na,nb & 01:38:59.41 & 14.208 & 12.6 & 50-900 & &   \\
\hline
\multicolumn{8}{c}{\textcolor{black}{bn170830135}} \\
\href{https://heasarc.gsfc.nasa.gov/FTP/fermi/data/gbm/bursts/2017/bn170830135/quicklook/glg_lc_all_bn170830135.gif}{GBM onboard} & n9,na & 03:14:01.95 & 0.512 & 4.7 & 47-291 & \multirow{4}{*}{long} & \href{https://gcn.gsfc.nasa.gov/other/525755646.fermi}{Notice}  \\
\FWBSB{} & n9,na &  03:14:01.15   &    12.416 &     10.9  &   4-50 & & \multirow{4}{*}{\href{https://github.com/santier14/FWBSB/tree/master/August_2017/170830/bn170830135-WBS}{Link}}  \\
\FWBSB{} & n2,n9,na,nb & 03:14:00.64  &     45.312     & 8.1   & 4-900 & &   \\
\FWBSB{} & n9,na & 03:14:01.03  &     6.272    &  15.0 & 50-900 & &   \\
\FWBSB{} & n2,n9,na & 03:14:01.16 & 6.144  & 9.0 & 50-100 & &   \\
\href{http://maxi.riken.jp/alert/novae/7995118832/7995118832.htm}{MAXI} & - & 03:15:29 & - & - & 4-10 &   & \href{https://gcn.gsfc.nasa.gov/gcn3/21761.gcn3}{Notice}  \\
\href{bb}{Astrosat-CZT} & - & 03:14:00 & - & - & 4-200 &   & \href{https://gcn.gsfc.nasa.gov/gcn3/21773.gcn3}{Notice}  \\
\hline
\multicolumn{8}{c}{\textcolor{black}{bn170830328}} \\
\href{https://heasarc.gsfc.nasa.gov/FTP/fermi/data/gbm/bursts/2017/bn170830328/quicklook/glg_lc_all_bn170830328.gif}{GBM onboard} & n3,n4 & 07:51:51.11 & 2.048 & 5.1 & 47-291 & \multirow{4}{*}{long} & \href{https://gcn.gsfc.nasa.gov/other/525772316.fermi}{Notice}  \\
\FWBSB{} & n3,n4 & 07:51:47.78  &     9.216    &  14.2  &  4-50 & & \multirow{4}{*}{\href{https://github.com/santier14/FWBSB/tree/master/August_2017/170830/bn170830328-WBS}{Link}}   \\
\FWBSB{} & n3,n4,n5,n8 & 07:51:48.80    &    8.832  & 15.7 & 4-900 & &   \\
\FWBSB{} & n1,n5,n8 & 07:51:48.30  & 8.832 & 6.2  &  50-100 & &   \\
\FWBSB{} & n5 & 07:51:48.94 & 7.808   & 10.9  & 50-900 & &   \\
\href{http://cgbm.calet.jp/cgbm_trigger/flight/1188114098/}{CALET} & - & 07:51:52.37 & 4.0 & 11.6 & 4-100 &   & \href{https://gcn.gsfc.nasa.gov/gcn3/21855.gcn3}{Notice}  \\
\href{http://www.ioffe.ru/LEA/kw/triggers/2017/kw20170830_28310.html}{Konus-WIND} & - & 07:51:50.16 & 80.509 & - & 50-200 &   & -  \\
\end{tabular}
\end{table*}

\begin{table*}
\contcaption{Results of the \FWBSB{} data analysis (August 2017, August 2018 and September 2018) compared with the \href{https://gcn.gsfc.nasa.gov/fermi_grbs.html}{onboard GBM analysis}. As an indication, other surveys that independently triggered on the same gamma-ray bursts as HXMT, Swift-BAT, Konus-Wind, Calet and MAXI are reported as well.}
\begin{tabular}{cccccccc}
Search & Detectors & \multicolumn{3}{c}{Trigger} & Energy  & GRB  & Ref  \\
& Involved & Time (UT) & Duration (s) & SNR (av) &  band (keV)  & classification &  \\
\multicolumn{8}{c}{\textcolor{black}{August 2017}} \\
\hline
\multicolumn{8}{c}{\textcolor{black}{bn170831179}} \\
\href{https://heasarc.gsfc.nasa.gov/FTP/fermi/data/gbm/bursts/2017/bn170831179/quicklook/glg_lc_all_bn170831179.gif}{GBM onboard} & n0,n3 & 04:18:11.13 & 2.048 & 4.9 & 47-291 & \multirow{5}{*}{long} & \href{https://gcn.gsfc.nasa.gov/other/525845896.fermi}{Notice}  \\
\FWBSB{} & n0,n3,n4,n5 & 04:18:09.22   & 58.88    & 49.7  & 4-50 & &  \multirow{4}{*}{\href{https://github.com/santier14/FWBSB/tree/master/August_2017/170831/bn170831179-WBS}{Link}} \\
\FWBSB{} & n0,n1,n3 & 04:18:03.84    &   64.896 & 49.6 & 4-900 & &   \\
\FWBSB{} & n0,n1,n3,n4,n5 &  04:18:14.08  &    51.328     & 26.4  & 50-100 & &   \\ 
\FWBSB{} & n0,n1,n3 & 04:18:14.34    &  77.952  & 23.4 & 50-900 & &   \\ 
\href{http://www.ioffe.ru/LEA/kw/triggers/2017/kw20170831_15512.html}{Konus-WIND} & - & 04:18:32.76 & 48.859 & - & 50-200 &   & -  \\
\hline
\hline
\multicolumn{8}{c}{\textcolor{black}{August 2018}} \\
\hline
\hline
\multicolumn{8}{c}{\textcolor{black}{bn180809485}} \\
\href{https://heasarc.gsfc.nasa.gov/FTP/fermi/data/gbm/bursts/2018/bn180809485/quicklook/glg_lc_all_bn180809485.gif}{GBM onboard} & n9,nb & 11:38:12.08 & 1.024 & 6.0 & 47-291 & \multirow{5}{*}{long} & \href{https://gcn.gsfc.nasa.gov/other/555507497.fermi}{Notice}  \\
\FWBSB{}  & n6,na,nb & 11:38:09.98  &  17.408  & 11.8 & 4-50  & & \multirow{4}{*}{\href{https://github.com/santier14/FWBSB/tree/master/August-September_2018/180809/bn180809485-WBS}{Link}}  \\
\FWBSB{}  & n6,n9 & 11:38:06.92   &  14.976 &  11.8 &  50-100 &   \\
\FWBSB{}  & n6,n7,n9,na & 11:38:08.97  & 24.704   & 13.4 & 50-900  & &   \\
\FWBSB{}  & n6,n9,nb & 11:38:08.96 &   19.328  & 18.0 & 4-900  & &   \\
\href{http://maxi.riken.jp/alert/novae/8339385460/8339385460.htm}{MAXI} & - & 11:38:25 & - & - & 2-4 &   & \href{https://gcn.gsfc.nasa.gov/other/339385460.maxi}{Notice}  \\
\hline
\multicolumn{8}{c}{\textcolor{black}{bn180810278}} \\
\href{https://heasarc.gsfc.nasa.gov/FTP/fermi/data/gbm/bursts/2018/bn180810278/quicklook/glg_lc_all_bn180810278.gif}{GBM onboard} & n8,nb & 06:40:46.74 & 2.048 & 5.1 & 47-291 & \multirow{5}{*}{long} & \href{https://gcn.gsfc.nasa.gov/other/555576051.fermi}{Notice}  \\
\FWBSB{}  & n7,n8,nb & 06:40:44.35 &  8.064  & 10.3 & 4-50  & &  \multirow{4}{*}{\href{https://github.com/santier14/FWBSB/tree/master/August-September_2018/180810/bn180810278-WBS}{Link}} \\
\FWBSB{}  & n8,nb & 06:40:45.00 &   14.848  & 9.2 & 50-100  & &   \\
\FWBSB{}  & nb & 06:40:44.10 &  15.744 &  11.1 & 50-900  & &   \\
\FWBSB{}  & n7,n8,nb & 06:40:44.35  &  15.36  & 13.4 & 4-900  & &   \\
\hline
\multicolumn{8}{c}{\textcolor{black}{bn180812349}} \\
\href{https://heasarc.gsfc.nasa.gov/FTP/fermi/data/gbm/bursts/2018/bn180812349/quicklook/glg_lc_all_bn180812349.gif}{GBM onboard} & n1,n2 & 08:22:30.31 & 0.128 & 5.1 & 47-291 & \multirow{2}{*}{long} & \href{https://gcn.gsfc.nasa.gov/other/555754955.fermi}{Notice}  \\
\FWBSB{}  & n0 & 08:22:29.70 &  21.76   & 7.3 & 4-900  & &  \multirow{1}{*}{\href{https://github.com/santier14/FWBSB/tree/master/August-September_2018/180812/bn180812349-WBS}{Link}} \\
\href{https://gcn.gsfc.nasa.gov/other/852903.swift}{Swif-BAT} & - & 08:22:30.23 & 2.048 & 11.2 & 25-100 &   & \href{https://gcn.gsfc.nasa.gov/notices_s/852903/BA/}{Notice}  \\
\hline
\multicolumn{8}{c}{\textcolor{black}{bn180814505}} \\
\href{https://heasarc.gsfc.nasa.gov/FTP/fermi/data/gbm/bursts/2018/bn180814505/quicklook/glg_lc_all_bn180814505.gif}{GBM onboard} & n6,n9,nb & 12:06:54.07 & 2.048 & 4.7 & 47-291 & \multirow{3}{*}{long} & \href{https://gcn.gsfc.nasa.gov/other/555941219.fermi}{Notice}  \\
\FWBSB{}  & n6 & 12:07:35.62  &  11.008 & 6.3 & 4-900  & & \multirow{2}{*}{\href{https://github.com/santier14/FWBSB/tree/master/August-September_2018/180814/bn180814505-WBS}{Link}}   \\
\FWBSB{}  & n9 & 12:06:51.00  &  3.328 & 6.4 & 4-900  & &   \\
\hline
\multicolumn{8}{c}{\textcolor{black}{bn180816088}} \\
\href{https://heasarc.gsfc.nasa.gov/FTP/fermi/data/gbm/bursts/2018/bn180816088/quicklook/glg_lc_all_bn180816088.gif}{GBM onboard} & n0,n1,n2 & 02:07:18.91 & 1.024 & 4.9 & 47-291 & \multirow{5}{*}{long} & \href{https://gcn.gsfc.nasa.gov/other/556078043.fermi}{Notice}  \\
\FWBSB{}  & n0,n1,n2,n5 & 02:07:16.93 &  42.496  & 60.3 & 4-50  & &  \multirow{4}{*}{\href{https://github.com/santier14/FWBSB/tree/master/August-September_2018/180816/bn180816088-WBS}{Link}}  \\
\FWBSB{}  & n0,n1,n5 & 02:07:17.95 &   43.008 & 26.3 & 50-100  & &   \\
\FWBSB{}  & n0, n1, n3, n5  & 02:07:17.83 &  40.192 &  25.9 & 50-900  & &   \\
\FWBSB{}  & n0,n1,n2,n3,n5 & 02:07:16.93  &  41.472  & 63.7 & 4-900  & &   \\
\href{http://www.hxmt.org/images/GRB/HEB180816088_lc.jpg}{HXMT} & - & 02:07:18.91 & 36.58 & - & 80-800 &   & -  \\
\href{http://www.ioffe.ru/LEA/kw/triggers/2018/kw20180816_07654.html}{Konus-WIND} & - & 02:07:34.83 & 31.261 & - & 50-200 &   & -  \\
\hline
\multicolumn{8}{c}{\textcolor{black}{bn180816930}} \\
\href{https://heasarc.gsfc.nasa.gov/FTP/fermi/data/gbm/bursts/2018/bn180816930/quicklook/glg_lc_all_bn180816930.gif}{GBM onboard} & n3,n4 & 22:19:53.32 & 1.024 & 5.4 & 47-291 &  \multirow{5}{*}{long} & \href{https://gcn.gsfc.nasa.gov/other/556078043.fermi}{Notice}  \\
\FWBSB{}  & n3,n4 & 22:19:51.23 &  9.984  & 11.1 & 4-50  & &  \multirow{4}{*}{\href{https://github.com/santier14/FWBSB/tree/master/August-September_2018/180816/bn180816930-WBS}{Link}} \\
\FWBSB{}  & n3,n4 & 22:19:51.37 &   9.6  & 9.0 & 50-100  & &   \\
\FWBSB{} & n3,n4 & 22:19:51.49 &  9.728 &  9.6  & 50-900  & &   \\
\FWBSB{}  & n3,n4,n5 & 22:19:51.49 &  10.368 &  11.9 & 4-900  & &   \\
\hline
\multicolumn{8}{c}{\textcolor{black}{bn180818179}} \\
\href{https://heasarc.gsfc.nasa.gov/FTP/fermi/data/gbm/bursts/2018/bn180818179/quicklook/glg_lc_all_bn180818179.gif}{GBM onboard} & n2,n5 & 04:17:58.30 & 0.128 & 5.7 & 47-291 & \multirow{2}{*}{long}& \href{https://gcn.gsfc.nasa.gov/other/556258682.fermi}{Notice}  \\
\FWBSB{}  & n2 & 04:17:56.67 & 0.768 & 5.1 & 50-900  & & \multirow{1}{*}{\href{https://github.com/santier14/FWBSB/tree/master/August-September_2018/180818/bn180818179-WBS}{Link}}  \\    
\hline
\multicolumn{8}{c}{\textcolor{black}{bn180818520}} \\
\href{https://heasarc.gsfc.nasa.gov/FTP/fermi/data/gbm/bursts/2018/bn180818520/quicklook/glg_lc_all_bn180818520.gif}{GBM onboard} & n2,n5 & 12:28:58.24 & 0.128 & 5.7 & 47-291 & \multirow{3}{*}{long} & \href{https://gcn.gsfc.nasa.gov/other/556258682.fermi}{Notice}  \\
\FWBSB{}  & n0,n1,n2 & 12:28:44.74 &  30.848  & 6.6 & 4-50  & & \multirow{4}{*}{\href{https://github.com/santier14/FWBSB/tree/master/August-September_2018/180818/bn180818520-WBS}{Link}}  \\
\FWBSB{}  & n1,n5 & 12:28:47.30 &  21.12 &  6.6 & 4-900  & &   \\
\FWBSB{}  & n2,n5 & 12:30:37.63 &  19.712 &  6.4 & 4-50  & &   \\
\FWBSB{}  & n2,n5 & 12:30:38.66 &  22.272  & 6.4 & 4-900  & &  \\
\href{https://gcn.gsfc.nasa.gov/other/853882.swift}{Swif-BAT} & - & 12:30:18.55 & 128 & 10.2 & 15-50 &   & \href{https://gcn.gsfc.nasa.gov/notices_s/853882/BA/}{Notice}  \\
\end{tabular}
\end{table*}
\begin{table*}
\contcaption{Results of the \FWBSB{} data analysis (August 2017, August 2018 and September 2018) compared with the \href{https://gcn.gsfc.nasa.gov/fermi_grbs.html}{onboard GBM analysis}. As an indication, other surveys that independently triggered on the same gamma-ray bursts as HXMT, Swift-BAT, Konus-Wind, Calet and MAXI are reported as well.}
\begin{tabular}{cccccccc}
Search & Detectors & \multicolumn{3}{c}{Trigger} & Energy  & GRB  & Ref  \\
& Involved & Time (UT) & Duration (s) & SNR (av) &  band (keV)  & classification &  \\
\multicolumn{8}{c}{\textcolor{black}{August 2017}} \\
\hline
\multicolumn{8}{c}{\textcolor{black}{bn180821653}} \\
\href{https://heasarc.gsfc.nasa.gov/FTP/fermi/data/gbm/bursts/2018/bn180821653/quicklook/glg_lc_all_bn180821653.gif}{GBM onboard} & n3,n6 & 15:40:39.22 & 4.096 & 4.7 & 47-291 & \multirow{2}{*}{long} & \href{https://gcn.gsfc.nasa.gov/other/556558844.fermi}{Notice}  \\
\FWBSB{}  & n3 & 15:41:32.04 &  0.768 &  36.3 & 4-900  & & \multirow{1}{*}{\href{https://github.com/santier14/FWBSB/tree/master/August-September_2018/180821/bn180821653-WBS}{Link}}  \\
\hline
\multicolumn{8}{c}{\textcolor{black}{bn180822423}} \\
\href{https://heasarc.gsfc.nasa.gov/FTP/fermi/data/gbm/bursts/2018/bn180822423/quicklook/glg_lc_all_bn180822423.gif}{GBM onboard} & n3,n4,n7 & 10:08:27.90 & 0.256 & 5.1 & 47-291 & \multirow{5}{*}{long} & \href{https://gcn.gsfc.nasa.gov/other/556625312.fermi}{Notice}  \\
\FWBSB{}  & n3,n4 &  10:08:26.82  & 8.704 & 16.8 &    4-50  & &  \multirow{4}{*}{\href{https://github.com/santier14/FWBSB/tree/master/August-September_2018/180822/bn180822423-WBS}{Link}}  \\
\FWBSB{} & n3,n4,n6,n7 & 10:08:26.44 &   8.448  &  12.1 &  50-100  & &   \\
\FWBSB{}  & n3,n4,n5,n6,n7 & 10:08:26.56  & 9.344  & 10.5 & 50-900  & &   \\
\FWBSB{} & n3,n4,n6 & 10:08:26.56 & 8.448 & 17.6  & 4-900  & &   \\
\hline
\multicolumn{8}{c}{\textcolor{black}{bn180822562}} \\
\href{https://heasarc.gsfc.nasa.gov/FTP/fermi/data/gbm/bursts/2018/bn180822562/quicklook/glg_lc_all_bn180822562.gif}{GBM onboard} & n0,n1 & 13:28:34.50 & 0.256 & 5.1 & 47-291 & \multirow{5}{*}{long} & \href{https://gcn.gsfc.nasa.gov/other/556637319.fermi}{Notice}  \\
\FWBSB{}  & n0,n1 & 13:28:35.01 & 2.816 & 23.0  & 4-50  & & \multirow{4}{*}{\href{https://github.com/santier14/FWBSB/tree/master/August-September_2018/180822/bn180822562-WBS}{Link}}    \\
\FWBSB{}  & n0,n3 & 13:28:34.00 & 3.968 &   13.6 & 50-100  & &   \\
\FWBSB{}  & n0,n1,n3 & 13:28:33.48  & 4.48 & 16.5 & 50-900  & &   \\
\FWBSB{}  & n0,n1,n3 &  13:28:33.22  & 4.864  &  21.5 &   4-900  & &   \\
\href{http://www.hxmt.org/images/GRB/HEB180822561_lc.jpg}{HXMT} & - & 13:28:34.05 & 3.677 & - & 80-800 &   & -  \\
\href{http://www.ioffe.ru/LEA/kw/triggers/2018/kw20180822_48514.html}{Konus-WIND} & - & 13:28:34.47 & 128.9 & - & 50-200 &   & -  \\
\hline
\multicolumn{8}{c}{\textcolor{black}{bn180823442}} \\
\href{https://heasarc.gsfc.nasa.gov/FTP/fermi/data/gbm/bursts/2018/bn180823442/quicklook/glg_lc_all_bn180823442.gif}{GBM onboard} & n1,n9 & 10:36:39.23 & 1.024 & 5.5 & 47-291  & \multirow{3}{*}{long} & \href{https://gcn.gsfc.nasa.gov/other/556713403.fermi}{Notice} \\
\FWBSB{}  & n6,n9 &  10:36:35.14  & 20.9 & 9.6 & 50-900  & & \multirow{2}{*}{\href{https://github.com/santier14/FWBSB/tree/master/August-September_2018/180823/bn180823442-WBS}{Link}}   \\
\FWBSB{} & na,nb &  10:36:36.42  & 7.7   &  8.4 &   4-900  & &   \\ 
\hline
\multicolumn{8}{c}{\textcolor{black}{bn180826055}} \\
\href{https://heasarc.gsfc.nasa.gov/FTP/fermi/data/gbm/bursts/2018/bn180826055/quicklook/glg_lc_all_bn180826055.gif}{GBM onboard} & n1,n9 & 01:19:15.54 & 2.048 & 4.5 & 47-291  & \multirow{4}{*}{long} & \href{https://gcn.gsfc.nasa.gov/other/556939160.fermi}{Notice}  \\
\FWBSB{}  & n1,n6 & 01:20:12.23  & 21.76  & 7.85   &  4-50  & & \multirow{3}{*}{\href{https://github.com/santier14/FWBSB/tree/master/August-September_2018/180826/bn180826055-WBS}{Link}}    \\
\FWBSB{}  & n0,n4,na &  01:20:06.59  & 34.816  &   14.1 & 50-900  & &   \\
\FWBSB{}  & n1,n2,n6,nb & 01:20:14.14 & 22.784 &  15.45 & 4-900  & &   \\
\href{http://www.ioffe.ru/LEA/kw/triggers/2018/kw20180826_04829.html}{Konus-WIND} & - & 01:20:29.84 & 211.78 & - & 50-200 &   & -  \\
\hline
\multicolumn{8}{c}{\textcolor{black}{bn180828790 - NOT A GRB}} \\
\href{https://heasarc.gsfc.nasa.gov/FTP/fermi/data/gbm/bursts/2018/bn180828790/quicklook/glg_lc_all_bn180828790.gif}{GBM onboard} & n9,na,nb & 18:57:39 & 0.519 & 4.9 & 47-291  & \multirow{4}{*}{-} & \href{https://gcn.gsfc.nasa.gov/other/557175451.fermi}{Notice}  \\
\FWBSB{}   & \multicolumn{7}{c}{No data available} \\ 
\href{https://gcn.gsfc.nasa.gov/other/856977.swift}{Swif-BAT} & - & 18:57:22.49 & 1.024 & 120.68 & 50-350 &   & \href{https://gcn.gsfc.nasa.gov/notices_s/856977/BA/}{Notice}  \\
\href{http://www.hxmt.org/images/GRB/HEB180828789_lc.jpg}{HXMT} & - & 18:57:26.58 & 7.607 & - & 80-800 &   & -  \\
\href{http://www.ioffe.ru/LEA/kw/triggers/2018/kw20180828_68250.html}{Konus-WIND} & - & 18:57:30.02 & 8.334 & - & 50-200 &   & -  \\
\hline
\hline
\multicolumn{8}{c}{\textcolor{black}{September 2018}} \\
\hline
\hline
\multicolumn{8}{c}{\textcolor{black}{bn180905400}} \\
\href{https://heasarc.gsfc.nasa.gov/FTP/fermi/data/gbm/bursts/2018/bn180905400/quicklook/glg_lc_all_bn180905400.gif}{GBM onboard} & n9,nb & 09:36:09.67 & 1.024 & 5.9 & 47-291 & \multirow{5}{*}{long} & \href{https://gcn.gsfc.nasa.gov/other/557832974.fermi}{Notice}   \\
\FWBSB{} & n9,na & 09:36:10.18  & 6.016  & 8.3 & 4-50 & & \multirow{4}{*}{\href{https://github.com/santier14/FWBSB/tree/master/August-September_2018/180905/bn180905400-WBS}{Link}}   \\ 
\FWBSB{} & n9,na & 09:36:08.77 & 6.016 & 10.8 & 4-900 & &   \\
\FWBSB{} & n9,na & 09:36:08.52 &  5.12 & 8.9  & 50-100 & &   \\
\FWBSB{} & n1,n9,na & 09:36:08.51  & 4.608 & 7.5   & 50-900 & &   \\   
\hline
\multicolumn{8}{c}{\textcolor{black}{bn180906597}} \\
\href{https://heasarc.gsfc.nasa.gov/FTP/fermi/data/gbm/bursts/2018/bn180906597/quicklook/glg_lc_all_bn180906597.gif}{GBM onboard} & n1,n2 & 14:19:51.78 & 0.256 & 4.5 & 47-291 & \multirow{5}{*}{long} & \href{https://gcn.gsfc.nasa.gov/other/557936396.fermi}{Notice}  \\
\FWBSB{} & n0,n1,n2,n5 & 14:19:52.26 & 14.72 & 12.1    & 4-50 & & \multirow{4}{*}{\href{https://github.com/santier14/FWBSB/tree/master/August-September_2018/180906/bn180906597-WBS}{Link}}    \\
\FWBSB{} & n1,n2,n5 & 14:19:50.21 & 12.288  & 17.3  & 4-900 & &   \\   
\FWBSB{} & n5 & 14:19:50.35  &  9.216 & 12.7 & 50-100 & &   \\   
\FWBSB{} & n0,n1,n2,n5 & 14:19:49.83 & 9.216 & 12.6 & 50-900 & &   \\  
\hline
\multicolumn{8}{c}{\textcolor{black}{bn180906759}} \\
\href{https://heasarc.gsfc.nasa.gov/FTP/fermi/data/gbm/bursts/2018/bn180906759/quicklook/glg_lc_all_bn180906759.gif}{GBM onboard} & n6,nb & 18:12:25.94 & 0.512 & 5.1 & 47-291 &  \multirow{5}{*}{long} & \href{https://gcn.gsfc.nasa.gov/other/557950350.fermi}{Notice}  \\
\FWBSB{} & n6,n7,n8,n9,nb & 18:12:25.02  & 11.136 & 13.8   & 4-50 & & \multirow{4}{*}{\href{https://github.com/santier14/FWBSB/tree/master/August-September_2018/180906/bn180906759-WBS}{Link}}   \\   
\FWBSB{} & n6,n7,n8,n9,nb & 18:12:25.15 & 10.112 & 22.7  & 4-900 & &   \\ 
\FWBSB{} & n6,n7,n8,n9,nb & 18:12:25.29 & 7.168 & 15.3   & 50-100 & &   \\
\FWBSB{} & n6,n8,n9,nb & 18:12:25.17 & 6.4 & 19.3  & 50-900 & &   \\ 
\href{http://www.ioffe.ru/LEA/kw/triggers/2018/kw20180828_68250.html}{Konus-WIND} & - & 18:12:27.92 & 6.613 & - & 50-200 &   & -  \\
\end{tabular}
\end{table*}
\begin{table*}
\contcaption{Results of the \FWBSB{} data analysis (August 2017, August 2018 and September 2018) compared with the \href{https://gcn.gsfc.nasa.gov/fermi_grbs.html}{onboard GBM analysis}. As an indication, other surveys that independently triggered on the same gamma-ray bursts as HXMT, Swift-BAT, Konus-Wind, Calet and MAXI are reported as well.}
\begin{tabular}{cccccccc}
Search & Detectors & \multicolumn{3}{c}{Trigger} & Energy  & GRB  & Ref  \\
& Involved & Time (UT) & Duration (s) & SNR (av) &  band (keV)  & classification &  \\
\multicolumn{8}{c}{\textcolor{black}{August 2017}} \\
\hline
\multicolumn{8}{c}{\textcolor{black}{bn180906988}} \\
\href{https://heasarc.gsfc.nasa.gov/FTP/fermi/data/gbm/bursts/2018/bn180906988/quicklook/glg_lc_all_bn180906988.gif}{GBM onboard} & n5,n6 & 23:42:34.16 & 0.256 & 5.2 & 47-291 & \multirow{3}{*}{long} & \href{https://gcn.gsfc.nasa.gov/other/557970159.fermi}{Notice}  \\
\FWBSB{} & n6,n7 & 23:42:33.66  & 3.712 & 12.5   & 4-50 & & \multirow{2}{*}{\href{https://github.com/santier14/FWBSB/tree/master/August-September_2018/180906/bn180906988-WBS}{Link}}  \\ 
\FWBSB{} & n6,n7 & 23:42:33.66 & 3.84 & 13.5 & 4-900 & &   \\
\hline
\end{tabular} 
\end{table*}

\end{document}